%% file: main.tex
\documentclass[conference,10pt,onecolumn]{IEEEtran}
\usepackage{blindtext, graphicx}{}
\usepackage{amsmath,amsfonts,amssymb,bbm}
\usepackage{mathtools}

\usepackage{amsthm}

\DeclareMathOperator*{\argmin}{arg\,min}

\usepackage[T1]{fontenc}

\usepackage{epstopdf}
\usepackage{epsfig}
\graphicspath{{images/}}
\usepackage{dsfont}
\usepackage{psfrag,bbm,graphicx}
\usepackage{comment,balance}
\usepackage{hyperref}
\usepackage{epstopdf}
\usepackage{epsfig}
\usepackage{multicol}
\usepackage{multirow, cite}
\usepackage{amsfonts}
\usepackage{color}
\usepackage[ruled,vlined]{algorithm2e}

\usepackage{lmodern}
\usepackage{amsmath,array}
\usepackage{easybmat}
\DeclarePairedDelimiter\ceil{\lceil}{\rceil}
\DeclarePairedDelimiter\floor{\lfloor}{\rfloor}

\hypersetup{
	colorlinks=true,
	linkcolor=blue,
	filecolor=magenta,      
	urlcolor=cyan,
}

\ifCLASSINFOpdf

\else

\fi

\renewcommand{\thepage}{\arabic{page}}
\renewcommand{\thesection}{\Roman{section}}
\renewcommand{\theequation}{\arabic{equation}}
\renewcommand{\thefigure}{\arabic{figure}}
\renewcommand{\thetable}{\arabic{table}}

\newtheorem{theorem}{Theorem}
\newtheorem*{theorem*}{Theorem}
\newtheorem{lemma}{Lemma}
\newtheorem*{lemma*}{Lemma}

\theoremstyle{claim}

\newtheorem{claim}{Claim}
\newtheorem*{claim*}{Claim}
\newtheorem{example}{Example}
\newtheorem{corollary}[theorem]{Corollary}

\theoremstyle{definition}
\newtheorem{definition}{Definition}[section]

\newtheorem{remark}[theorem]{Remark}
\newcommand{\remove}[1]{}

\title{On Gradient Coding with Partial Recovery}
\begin{document}

\author{%
  \IEEEauthorblockN{Sahasrajit Sarmasarkar\IEEEauthorrefmark{1},
  V. Lalitha\IEEEauthorrefmark{2} 
  and Nikhil Karamchandani\IEEEauthorrefmark{1}}\\
  \IEEEauthorblockA{\IEEEauthorrefmark{1}%
                    Indian Institute of Technology, Bombay\\
                   Email: sahasrajit1998@gmail.com, nikhilk@ee.iitb.ac.in}
   
    \IEEEauthorblockA{\IEEEauthorrefmark{2}%
    International Institute of Information Technology, Hyderabad\\
                   Email: lalitha.v@iiit.ac.in}
                   
  \thanks{This paper was presented in part at the 2021 International Symposium on Information Theory (ISIT) \cite{sarmasarkar2021isit}.}
}
\maketitle
\begin{abstract}
    We consider a generalization of the gradient coding framework where a dataset is divided across $n$ workers and each worker transmits to a master node one or more linear combinations of the gradients over its assigned data subsets. Unlike the conventional framework which requires the master node to recover the sum of the gradients over all the data subsets in the presence of straggler workers, we relax the goal to computing the sum of at least some $\alpha$ fraction of the gradients. We begin by deriving a lower bound on the computation load of any scheme and also propose two strategies which achieve this lower bound, albeit at the cost of high communication load and a number of data partitions which can be polynomial in $n$. We then propose schemes based on cyclic assignment which utilize $n$ data partitions and have a lower communication load. When each worker transmits a single linear combination, we prove lower bounds on the computation load of any scheme using $n$ data partitions. Finally, we describe a class of schemes which achieve different intermediate operating points for the computation and communication load and provide simulation results to demonstrate the empirical performance of our schemes.
%
 %
\end{abstract}

\input{intro.tex}
\input{problem_formulation}
\input{lower_bound1}

\input{comp_min_scheme.tex}

\input{comp_min_scheme_improve}

\input{cyclic_scheme.tex}

\input{scheme_low_comp_comm_1.tex}

\input{simulations}

\input{discussion.tex}

\appendices

{Note that for the proofs in the upcoming appendix, we denote $[n]$ as the set of consecutive integers from 1 to $n$. Also we denote the remainder when $b$ divides $a$ by $a \% b$ or $a \mod \ b$}

\input{decr_claim_proof.tex}

\input{constr_lower_comm.tex}

\input{proof_cyclic_constr_1.tex}

\input{proof_cyclic_constr_2}

\input{proof_lower_bound_low_comp}

\input{proof_interim_points}


\bibliographystyle{IEEEtran}
\bibliography{refs}

\end{document}

%% file: intro.tex
\section{Introduction}
In a distributed computing framework, a job is divided into multiple parallel tasks, which are computed on different servers, and the job is finished when all the tasks are complete. In this framework, a subset of workers can be arbitrarily slow as compared to the rest of the workers.
These subset of workers are referred to as stragglers. Since the slowest tasks determine the job execution time, they form a bottleneck to the efficient execution of the job. Recently, there has been an extensive amount of work to mitigate the effect of stragglers by introducing redundancy in the computed tasks using coding theoretic techniques; see for example \cite{li2020coded} for an extensive survey. The distributed computing applications for which codes have been designed include matrix-vector multiplication \cite{dutta2016short}, matrix-matrix multiplication \cite{lee2017speeding}, \cite{yu2017polynomial}, gradient computation  \cite{tandon2017gradient}, polynomial computation \cite{yu2019lagrange} {and coded convolution \cite{8006960}}. A fundamental trade-off between computation and communication cost was established in \cite{li2017fundamental}, for the case of the general distributed data shuffling problem.

\subsection{Gradient Coding}
\remove{\color{red}Please go over all references and make sure they are up to date and accurate in terms of their publication details.Comment SS- updated}

In various machine learning applications, a principal task is to compute the gradient sum  on large datasets. Hence, gradient sum computation is a natural application for distributed computing. Consider a dataset of $d$ points over which the gradient sum of a certain objective function needs to be computed. In the case of uncoded computing, the data set is divided into $n$ data subsets. Each worker computes a partial gradient on the data subset assigned to it and returns the results to the master node. The master computes the full gradient sum by combining the results. However, this scheme is not efficient when there are stragglers amongst the $n$ worker nodes. Towards addressing this issue, \textit{Gradient Coding} was proposed in \cite{tandon2017gradient}, which ensures efficient distributed gradient computation even in the presence of stragglers by utilizing coding-theoretic techniques. For any scheme which is tolerant to $s$ stragglers, a lower bound of $s+1$ on the computation load per worker was derived. Optimal gradient coding schemes, which achieve the lower bound with equality, were provided based on fractional repetition and cyclic assignments of data subsets.

{The scheme based on cyclic assignment of data subsets in \cite{tandon2017gradient}, is based on a random coding argument. \remove{and hence the result is existential in nature}} Gradient coding schemes based on cyclic MDS codes over complex numbers and on Reed Solomon codes were designed in \cite{raviv2020gradient} and \cite{halbawi2018improving} respectively. When gradient sum computation can be formulated as a multivariate polynomial evaluation problem, the Lagrange coded computing scheme has been proposed in \cite{yu2019lagrange}.  Communication-efficient gradient coding was introduced in \cite{ye2018communication} where the master node has to recover a gradient sum vector and it proposes coding across the elements of the gradient vector to reduce the number of transmitted symbols. Multi-message communication based gradient codes allow for multiple messages to be transmitted from workers to the master in each round and have been studied in \cite{ozfatura2019speeding}, \cite{ozfatura2020straggler} which use this capability to utilize the work done by non-persistent stragglers. Heterogeneity-aware gradient coding was introduced in \cite{wang2019heterogeneity}, where in addition to stragglers, heterogeneous non-straggling workers have been considered. The problem of distributed linearly separable computation has been introduced in \cite{9614153}. Distributed linear transforms and gradient computation are special cases of this problem.
\subsection{Approximate Gradient Coding}
The above works consider the objective of exactly recovering the gradients sum in the presence of stragglers and as mentioned before, a fundamental converse argument in \cite{tandon2017gradient} finds that this requires the per worker computation load to scale linearly with the straggler tolerance level. Several works have found that for many practical distributed learning applications, it suffices to approximately recover the  gradient sum \cite{dutta2018slow, chen2017revisiting, bottou2010large,wang2019erasurehead,9081964,maity2019robust}. Gradient coding schemes which trade-off the computation load and the $\ell_2$ error between the actual full gradient and the computed full gradient, have been studied in \cite{raviv2020gradient, wang2019fundamental, 8849690, charles2017approximate}. In this work, we consider a different form of approximation which we term gradient coding with partial recovery, where the gradient computed at the master is required to be the sum of at least $\alpha$ fraction of the data subsets. This setting is closely related to the $K$-sync SGD \cite{dutta2018slow} and fastest-$k$ SGD \cite{DBLP:journals/corr/abs-2002-11005} studied in the literature, for which convergence guarantees and error analysis exists when the set of stragglers are chosen uniformly at random. Also such forms of approximate gradient recovery have found application in distributed learning algorithms \cite{dutta2018slow, chen2017revisiting}. A similar objective function was also studied recently in \cite{ozfatura2019distributed,9641837}, where a class of \textit{randomly circular shifted} codes were proposed for exploiting the partial computations performed by straggling workers, and  numerical simulations illustrated the advantages of allowing partial recovery. Finally, we would like to point out that while \cite{wang2019fundamental} studied approximate gradient coding in terms of $\ell_2$ error, their gradient code construction based on Batched Raptor codes can in fact be applied to the partial recovery framework being studied here as well. However, the guarantees are probabilistic in nature, where the randomization is on the set of stragglers whereas our focus here is on the deterministic worst-case setup as proposed for the original gradient coding problem \cite{tandon2017gradient}.

\subsection{Our Contributions}
We introduce gradient coding with partial recovery as a generalization of the standard gradient coding framework widely studied in the literature and make the following contributions towards understanding the fundamental communication-computation load tradeoff in this context. 

 \noindent (a) We begin by proving a lower bound in Theorem~\ref{final_lb} on the computation load at each worker for any scheme which is tolerant to $s$ stragglers and for any given partial recovery level $\alpha$. For $\alpha = 1$, this expression reduces to the known lower bound on computation load per worker for full gradient recovery \cite{tandon2017gradient}. \\
(b) Next, we provide two schemes which achieves this bound with equality in Theorems~\ref{compload_ub} and \ref{compub_load_tdesign}, the second one having  better communication load than the first but existing only for a subset of parameters. Though these schemes have minimum computation load, they have high communication load per worker in terms of the number of linear combinations of the assigned gradients each worker needs to communicate. Further, these schemes require partitioning the data into a large number of data subsets.\\ \remove{\color{red}(exponential in the number of workers)}
(c) In Section~\ref{cyclic_scheme_section}, we present a class of cyclic gradient codes with the number of data subsets being equal to the number of workers. These cyclic schemes have much lower communication load with every worker transmitting at most two linear combinations of the gradients of the data subsets assigned to it, but have a slightly higher computation load. The computation load and the communication load for these  $(\alpha,s)$-cyclic schemes for various parameters are summarised in Table~\ref{comm_comp_tradeoff_cyclic}. The second and third rows of the table indicate a tradeoff between the computation and communication load for these schemes. {Note that this is unlike the case of full gradient recovery with $\alpha = 1$ \cite{tandon2017gradient}, where cyclic schemes achieve the minimum computation load, with each worker transmitting just a single message}. In Section~\ref{sec:lowcomputation}, we also give a lower bound on the computation load per worker for \emph{any scheme} for which the number of data subsets is equal to the number of workers and where each worker transmits exactly one linear combination. \\
(d) Having discussed two complementary classes of schemes, one with low computation load and the other with the low communication load, in Section~\ref{compload_inter_points} we describe a whole class of $(\alpha,s)$ feasible GC schemes, which achieve different intermediate operating points for the computation and communication load. \\
(e) Finally, in Section~\ref{Sec:simulations}, we provide simulation results to compare the empirical performance of our proposed schemes with other approximate gradient coding schemes in the literature.
%
\begin{table}
    \centering
\begin{tabular}{|l|p{30mm}|p{25mm}|p{40mm}|} 
 \hline
 Theorem & Communication \newline cost & Computation \newline load & Constraint\\
 \hline
 \ref{cyclic_constr_comm_1}(Achievability) & 1 & $r/n$ & $(\beta \mod r)$ =0 \\
 \hline
 \ref{cyclic_scheme_lower_bound}(Lower bound) & 1 & $>r/n$ & $(\beta \mod r) \neq 0$ \\ 
 \hline
 \ref{cyclic_constr_comm_2}(Achievability) & 2 & $r/n$ & $r- (\beta \mod r) \leq n - \beta$ \\
 \hline
\end{tabular}
\vspace{1 em}
   \caption{Communication costs for cyclic schemes under various constraints.  $\beta = \ceil{\alpha.n}$ and $r=s+1+\beta-n$.}
    \label{comm_comp_tradeoff_cyclic}
\end{table}

%% file: problem_formulation.tex
\section{Problem Formulation}

Consider a dataset $D$ consisting of features-label pairs $\{(x_i,y_i)\}_{i=1}^{d}$ with each tuple $(x_i,y_i) \in \mathbb{R}^p\times \mathbb{R}$. Several machine learning problems wish to solve problems of the following form:

$$\beta ^{*} = \argmin_{\beta \in \mathbb{R}^{p}} \sum_{i=1}^{d} L(x_i,y_i;\beta) +\lambda R(\beta)$$

where $L(.)$ is a task-specific function and $R(.)$ is the regularisation function. Often this problem is solved using gradient-based iterative  approaches by computing the gradient at each step using the current value of the model $\beta^{(t)}$. Let $g^{(t)} := \sum\limits_{i=1}^{d} \nabla L(x_i,y_i:\beta^{(t)})$ be the gradient of the loss function computed at $t^{th}$ step and the model parameter is updated as $\beta^{(t+1)}=h_{R}(\beta^{(t)},g^{(t)})$ for some suitable mapping $h_R$. As the size $d$ of the dataset becomes large, the  computation of the gradient $g^{(t)}$ can become a bottleneck and one possible solution is to parallelize the computation by distributing the task across multiple workers.  

 \remove{gradient computation at each step is done by $d$ workers where  worker $i$ computes $g_i^{(t)} := \nabla L(x_i,y_i;\beta^{(t)})$. However, if the number of data points is large, failure of one worker may lead to a possible failure of the entire computation. We wish to design systems which would be tolerant to the computation failure of some workers.}

We consider a gradient coding framework with $n$ workers denoted by $W_1,W_2,...,W_n$ and a master node. The entire data set $D$ is divided into $k$ equal partitions $D_1, D_2,..,D_k$ and let $\{g_l\}$ denote the partial gradients\footnote{We drop the superscript $t$ in the gradient notation for convenience} over the data subsets $\{D_l\}$. Each worker $i$ computes $m \ge 1$ linear combinations of $\{g_l\}$  given by (coded partial gradient) $\tilde{g}^i =[\tilde{g}^i_1;\tilde{g}^i_2;...;\tilde{g}^i_m]$ with $\tilde{g}^i_j = \sum\limits_{l=1}^{k} A^i_{j,l} \cdot g_l$ for each $j \in [m]$, and transmits them to the master node. Let $A^i \in \mathbb{R}^{m\times k}$ denote the computation matrix corresponding to worker $i$ with its $(j,l)^{th}$ entry given by $A^i_{j,l}$. We define the communication load and computation load of the gradient coding scheme described above.
%
%
\begin{definition}
(Communication Load): For a gradient coding scheme specified by $\{A^i\}$, we define the communication load as $m$ where $m$ denotes the number of coded partial gradients transmitted by each worker.{\footnote{Note that this definition of communication cost differs from \cite{ye2018communication} which is defined as the number of dimensions of each transmitted vector.}}
\end{definition}
%

%
\begin{definition}
(Max. Computation Load per worker): For a gradient coding scheme with communication load $m$ and specified by $\{A^i\}$, we define the load per worker by $ l = \frac{1}{k}.\max\limits_{i \in [n]} |\bigcup\limits_{j \in [m]} \text{supp}(A^i_j)| $ where $\text{supp}(A^i_j)$ denotes the set of non-zero entries in the $j^{th}$ row of $A^i$.
\end{definition}
%
%
{The data subsets assigned to a worker $W_i$ is  $\{D_v: v \in \bigcup\limits_{j \in [m]} \text{supp}(A^i_j)\}$}. Note that we define the computation load relative to the total number of partitions $k$ of the entire data set. On the other hand, the communication load $m$ is not normalized since the size of each worker transmission is independent of the number of data subsets $k$. 

\remove{Clearly, $\bigcup\limits_{j \in [m]} \text{supp}(A^i_j)$ denotes the number of distinct gradients computed by worker $i$ and }

We will refer to a gradient coding scheme with $n$ workers, $k$ data subsets, communication load $m$, and maximum computation load per worker $l$ as an $(n,k,m,l)$ gradient coding (GC) scheme. In conventional gradient coding schemes, the goal of the master node is to recover the sum of the partial gradients $\{g_i\}$ over all the $k$ data subsets $\{D_i\}$ in the presence of straggler worker nodes. We now define a new framework in which the requirement for the master is relaxed to recovering the sum of a certain fraction of the partial gradients. %
\remove{
\begin{definition}
($\alpha$-recovery gradient coding schemes): A gradient coding scheme is $\alpha$ recovery coding schemes if the the master node should be able to compute $\sum\limits_{i \in I} g_i$ for some $I \subseteq [k]$ and $|I| \geq \alpha k$ whenever any $n-s$ workers are able to successfully communicate their results to the master node. 
\end{definition}
}
\begin{definition}
($(\alpha,s)$-feasible $(n,k,m,l)$ gradient coding (GC) schemes): For $\alpha \in (0, 1], 1 \le s \le n$, we call an $(n,k,m,l)$ gradient coding scheme as $(\alpha,s)$-feasible if the master node is able to compute $\sum\limits_{i \in I} g_i$ for some $I \subseteq [k]$, $|I| \geq \alpha k$ whenever any $n-s$ workers are able to successfully communicate their results to the master node. 
\end{definition}

Thus, if an $(n,k,m,l)$ GC is $(\alpha,s)$-feasible, then it can tolerate $s$ stragglers out of the $n$ workers. Also, note that for $\alpha =1$ the above definition reduces to that of conventional gradient codes. Finally, we will restrict attention to linear schemes here and thus for such a scheme, there must exist a vector $v \in \{0,1\}^k$ with $||v||_0\geq \alpha.k$ in the span of the rows of $\{A^i\}_{i \in I}$ for every $|I|\geq n-s$. 


Our goal in this work is to analyze the minimum communication load $(m)$ and computation load per worker $(l)$ for $(\alpha,s)$-feasible $(n,k,m,l)$ GC schemes. One naive strategy to create such a GC scheme is to select some $\alpha.k$ data partitions out of $D$ and then use a conventional (full) gradient coding scheme to recover the sum of gradients over the $\alpha.k$ data partitions while allowing for any set of $s$ workers to straggle. Such a scheme would have a communication load of $1$ and a lower bound of $\alpha(s+1) / n$ on the max. computation load per worker \cite{tandon2017gradient}. The intuition for this gain is that our schemes exploit the flexibility where we could recover the sum of any set of $\alpha.k$ gradients, which is absent in the naive approach. In this work, we will propose $(\alpha,s)$-feasible $(n,k,m,l)$ GC schemes that have far lower max. computation load per worker. 

%% file: lower_bound1.tex
\section{Lower bound on the computation load }
We begin by proving a lower bound on the computation load per worker $l$ for any $(\alpha,s)$-feasible $(n,k,m,l)$ GC scheme.
\begin{theorem}{\label{final_lb}}
    For any $(\alpha,s)$-feasible $(n,k,m,l)$ GC scheme and $y = \ceil{n.l}$, we have 
    \begin{equation}{\label{lb_equation}}
        \frac{{s \choose y}}{{n \choose y}} \leq 1 - \alpha .
    \end{equation}
    %
\end{theorem}
The inequality in \eqref{lb_equation} implies a lower bound on $y = \ceil{nl}$ and for a scheme which assigns the same load to each worker, $y$ denotes the average number of copies for each data subset stored across the $n$ workers. Note that while the above lower bound is dependent on the parameters $n$, $s$, and $\alpha$, it is independent of the number of data subsets $k$ and communication load $m$. Also, for $\alpha = 1$ which corresponds to the conventional gradient coding setup, the lower bound above reduces to $y\geq s+1$ as obtained in \cite[Theorem 1]{tandon2017gradient}.

To prove Theorem \ref{final_lb}, we derive an intermediate condition given in the following lemma.

\begin{lemma}{\label{init_lb}}
 Consider any $(\alpha,s)$-feasible $(n,k,m,l)$ GC scheme and let $y_i$ denote the number of distinct workers which are assigned the data subset $D_i$. Then, the following condition holds:
\begin{equation}{\label{lower_bound}}
    \sum_{i=1}^{k} {{n-y_i \choose n-s}} \leq {{n \choose s}}k (1-\alpha).
\end{equation}
\end{lemma}

\begin{proof}
Consider all possible subsets of size $s$ of the set of $n$ workers and denote these subsets by $\{S_j\}$ for $j \in [{{n \choose s}}]$. Now consider any data subset $D_i$ for some $i \in [k]$ and let $E_i$ denote the set of workers it is assigned to. From the statement of the lemma, we have $|E_i| = y_i$. From the definition of an $(\alpha,s)$-feasible $(n,k,m,l)$ GC scheme, we have that each subset of $(n-s)$ workers should have access to at least $\alpha$ fraction of the datasets, and thus for each subset $S_i$ of size $s$ there can be at most $k(1-\alpha)$ subsets $E_j$ such that $E_j \subseteq S_i$.
	
	For each $j \in [{{n \choose s}}]$, let $k_j = |\{E_i | i \in [k];E_i \subseteq S_j \}|$, whose sum we bound in the following argument. From the argument above, $\displaystyle \sum_{j \in [{{n \choose s}}]} k_j \leq {{n \choose s}} k(1-\alpha)$. On the other hand, each set $E_i$ is a subset of exactly ${{n-y_i \choose s-y_i}}$ subsets $S_j$ for  $j \in [{{n \choose s}}]$. Thus, we get $\displaystyle \sum_{i=1}^{k} {{n-y_i \choose n-s}} = \sum_{j \in [{{n \choose s}}]} k_j \leq {{n \choose s}}k (1-\alpha)$,  completing the proof.

\end{proof}

Now, we will use Lemma \ref{init_lb} \remove{ and Claim \ref{decr_claim} (stated after the proof of Theorem \ref{final_lb}) }to prove Theorem \ref{final_lb}.

\begin{proof}[Proof of Theorem~\ref{final_lb}]
   Consider any $(\alpha,s)$-feasible $(n,k,m,l)$ GC scheme and let $y_i$ denote the number of distinct workers which are assigned the data subset $D_i$. From the definition of the max. load per worker $l$, we have $\sum_{i\in [k]} {y_i}\leq n.k.l$ since each worker can be assigned at most $k.l$ data subsets. Furthermore, we have $\sum_{i=1}^{k} {{n-y_i \choose n-s}} \leq {{n \choose s}}k (1-\alpha)$ from Lemma~\ref{init_lb}.
   
\remove{
   We can show through some algebraic manipulations that $\sum_{i=1}^{k} {{n-y_i \choose n-s}}$ is the least when $\{y_i\}_{i \in [k]}$ differ by at most 1 and $\sum_i y_i=n.k.l$. Thus, we may lower bound $\sum_{i=1}^{k} {{n-y_i\choose n-s}}$ by $\sum_{i=1}^{k} {{n-y\choose n-s}}$ where $y=\ceil{\frac{n.k.l}{k}}$. Combining this inequality with Lemma~\ref{init_lb}, we get $k.{{n-y \choose n-s}} \leq k.{{n \choose s}}(1-\alpha)$ implying Theorem~\ref{final_lb} since $\frac{{{n-y \choose n-s}}}{{{n \choose s}}}=\frac{{s \choose y}}{{n \choose y}}$.   
}

    {Now define $b=\floor{\frac{\sum_{i=1}^{k} y_i}{k}}$ and $k_1=(b+1)k -\sum_{i=1}^{k} y_i$, thus from the claim \ref{decr_claim} below, $k_1{{n-b \choose n-s}}+(k-k_1){{n-b-1 \choose n-s}} \leq \sum_{i=1}^{k} {{n-y_i \choose n-s}}$ since $\sum_{i=1}^{k} (n-y_i) = k_1\times (n-b) + (k-k_1)\times(n-b-1)$ and $n-b-1=\floor{\frac{\sum_{i=1}^{n} (n-y_i)}{k}}$. The L.H.S is the smallest when $\sum_{i=1}^{k}{y_i} = n\times k \times l$ since $a$ increases with $\sum{y_i}$ and $k_1$ decreases with $\sum{y_i}$ when $a$ is constant. Thus, the inequality reduces to $k{{n-b-1 \choose n-s}}\leq  \sum_{i=1}^{k} {{n-y_i \choose n-s}}\leq {{n \choose s}}k (1-\alpha)$ where $a=\floor{{n.l}}$ because ${{n-b-1 \choose n-s}} \leq {{n-b \choose n-s}}$ which proves Theorem \ref{final_lb}.}
    
\end{proof}


\begin{claim}{\label{decr_claim}}
Consider any collection of $t$ positive integers $\{a_i\}_{1 \leq i \leq t}$. Define $a=\floor{\frac{\sum_{i=1}^{t} a_i}{t}}$ and let $t_1$ be the unique positive integer satisfying $\sum a_i = t_1.a+(t-t_1)(a+1)$. Then we have 
$\sum_{i=1}^{t}{{a_i \choose r}} \geq t_1.{{a \choose r}} + (t-t_1).{{a+1 \choose r}}$. 
\end{claim}

The proof of claim \ref{decr_claim} can be found in Appendix ~\ref{final_lb_details}.

%% file: comp_min_scheme.tex
\section{$(\alpha,s)$ feasible $(n,k,m,l)$ GC schemes with least computation load}\label{compload_ub_section}

The following theorem shows that lower bound on computation load in Theorem \ref{final_lb} is achievable, albeit at high communication cost. 

\begin{theorem}
\label{compload_ub}
For every $n,s,\alpha$ and $1 \le y \le n$ satisfying $\frac{{{s \choose y}}}{{{n \choose y}}} \leq 1- \alpha$, there exists an $(\alpha,s)$-feasible $(n,{{n \choose y}},{{n-1 \choose y-1}},\frac{y}{n})$ GC scheme.
\end{theorem}

\begin{proof}
 We divide the full dataset $D$ in to $k = {{n \choose y}}$ data subsets and index them by subsets of $[n]$ of size $y$, and for each $S \subset [n], S = \{i_1,i_2,\ldots,i_y\}$, let data subset $D_{S}$ be assigned to workers $W_{i_1}, W_{i_2},\ldots, W_{i_y}$.  Thus, each worker would be assigned ${{n-1 \choose y-1}}$ data subsets and the computation load per worker $l = \frac{ {{n-1 \choose y-1}} }{ {n \choose y} } = \frac{y}{n}$. Each worker would then directly compute and individually transmit the gradients for all the data subsets assigned to it, which results in a communication load $m = {{n-1 \choose y-1}}$. Next, we argue the correctness of this scheme.

\remove{
Thus the matrix $A^i$ for each worker would be of size ${{n-1 \choose y-1}} \times {{n \choose y}}$. Each column would be denoting a distinct set of $y$ workers and say we denote the $t^{th}$ data subset by the set $I_t$ of cardinality $y$ and each row denotes a distinct set of $y-1$ workers not containing worker $i$, thus the $j^{th}$ row in $A^i$ can be represented by a distinct set $I'_{i,j}$ of cardinality $y-1$ such that $i$ is not present in any of these sets.

Thus 
\begin{equation}
    A^i_{j,t} = 
    \begin{cases}
    1 & i \cup I'_{i,j} = I_t\\
    0 & \text{otherwise}
    \end{cases}
\end{equation}
$A^i_{j,t}=1$ if $i \cup I'_{i,j} = I_t$ else, 0. 
}
Under any set of $s$ stragglers, the number of data-parts which are not assigned to any worker other than these set of $s$ stragglers is given by ${{s \choose y}}$. Thus, the master node can obtain the sum of at least ${{n \choose y}}-{{s \choose y}}$ gradients which is at least $\alpha.{{n \choose y}} = \alpha k$ since $\frac{{{s \choose y}}}{{{n \choose y}}} \leq 1- \alpha$. Thus the above mentioned scheme is an $(\alpha,s)$-feasible GC scheme. 
\end{proof}

%% file: comp_min_scheme_improve.tex
\remove{
\subsection{Improvement in communication load when $n$ and $y$ are co-prime}\label{impr_comm_cost_section}}

We now show that the communication load can be slightly improved in some scenarios without incurring a penalty on the computation load per worker.
\begin{theorem}{\label{impr_comm_cost}}
For every $n,s,\alpha$ and $1 \le y \le n$ which is co-prime with $n$ and satisfies $\frac{{{s \choose y}}}{{{n \choose y}}} \leq 1- \alpha$, there exists an $(\alpha,s)$-feasible $(n,{{n \choose y}},1+\frac{y-1}{y}.{{n-1 \choose y-1}},\frac{y}{n})$ GC scheme.
\end{theorem}
%

\begin{proof}
We assign data subsets to different workers in the same way as described in the proof of Theorem~\ref{compload_ub} and thus the number of data partitions $k$ and the computation load per worker $l$ remain the same. Let $I_j \subset [n], |I_j| = y$ denote the indices of the $y$  workers to whom data subset $D_j$ is assigned. For each data subset $D_j$, we choose a worker $W_{t_j}$ from amongst the workers that data subset $D_j$ is assigned to, i.e., ${t_j} \in I_j$ $\forall j \in [{{n \choose y}}]$. This is done while ensuring that the process is balanced, i.e., each worker is chosen exactly the same number of times and thus we have $\forall \ i \in [n]$, $|B_i| = |\{j:  j \in [{{n \choose y}}]\text{ s.t } i = t_j\}|= {{n-1 \choose y-1}} / y$. Such an allocation is possible whenever $y$ is co-prime with $n$ and the details are provided in Appendix~\ref{proof_details_improved_comm}. Next, each worker $W_i$ transmits to the master node the sum of all the gradients assigned to it and in addition, individually transmits the gradients corresponding to all the data subsets assigned to it except those in $B_i$. Thus the communication load of this scheme is given by $1+ \frac{y-1}{y} {{n-1 \choose y-1}}$.
We now describe the decoding procedure at the master node and argue the correctness of the scheme in the presence of at most $s$ stragglers. Denote the set of non-straggler worker nodes by $I \subseteq [n]$ with $|I| \ge n - s$. Since the data subset assignment to the workers is identical to the one used in the proof of Theorem~\ref{compload_ub}, we know that the number of gradients which are computed by at least one worker in $I$ is greater than $\alpha k = \alpha {n\choose y}$. Thus to prove that the scheme is an $(\alpha,s)$-feasible GC, it suffices to show that using the transmissions from the non-straggling worker nodes, the master node can recover the sum of the gradients corresponding to all data subsets assigned to them. 

Recall that each non-straggler worker node in $I$ transmits the sum of all its computed gradients in addition to some individual gradients. The master node adds up the sum transmissions from all nodes in $I$ and then uses the individual gradient transmissions to suitably adjust the coefficients so that the sum of all the involved gradients can be recovered. Let $D_{1,I}$ denote the collection of data subsets which are assigned to exactly $1$ worker amongst the non-straggling workers $I$. Clearly, the gradient of each such data subset in $D_{1,I}$ would have its coefficient as $1$ in the above sum at the master node. Now consider the gradients of those data subsets which appeared more than once in the sum. Each such data subset must have been assigned to more than one worker in $I$ and thus at least one worker in $I$ would be directly transmitting the gradient of that data subset as per the scheme designed above. Thus, the master node can subtract an appropriate multiple of any such gradient from the sum calculated above and we can thus recover the sum of the gradients corresponding to all data subsets assigned to the non-straggling workers $I$.
\end{proof}

{
\begin{example}
  An example for $n=5$, $\alpha=7/10$ and $s=3$ is shown below as described above in proof of Thm~\ref{impr_comm_cost}. The smallest $y$ satisfying $\frac{{{s \choose y}}}{{{n \choose y}}} \leq 1-\alpha$ can be shown to 2. Since $n$ and $y$ are co-prime we can achieve a communication load of $1+\frac{y-1}{y}{{n-1 \choose y-1}}=3$.  The assignment of different data subsets to various workers is given in Table \ref{tab:example3}. Recall that under this scheme each worker $W_i$ transmits the sum of the gradients of data subsets it is assigned and individually transmits gradients corresponding to those data subsets except those in $B_i$. For each worker $W_i$, the data subsets assigned to it which belong to $B_i$ has been denoted by $1 \times$ and the data subsets assigned to it but don't belong in $B_i$ has been denoted by $1 \checkmark$.
  
  
  For example, worker $W_1$ transmits the sum of the gradients of the data subsets $D_1$, $D_2$, $D_3$ and $D_4$ and individually the gradients of the data subsets corresponding to data subsets $D_3$ and $D_4$.  For example if workers $W_3$, $W_4$ and $W_5$ straggle, the master can still compute the sum of gradients of subsets $D_1$ to $D_7$, $D_4$ , $D_6$, $D_7$ and $D_1$ using the transmissions by the workers $W_1$ and $W_2$. The master can compute the sum of the sum of the gradients transmitted by the workers $W_1$ and $W_2$ and subtract the gradient of the data subset $D_1$ which is transmitted by worker $W_2$. 
  
  Note that the scheme which just assigns only $\alpha$ fraction of data sets to the workers can be shown to have a lower bound on the max computation load per worker to be $\frac{\alpha(s+1)}{n}=0.56$.  \remove{Also the cyclic scheme as in \cite{tandon2017gradient} which calculates the sum of all the gradients must also assign exactly 4 data subsets to every worker.} Our scheme has a max. computation load per worker to be $\frac{2}{5}$ which is lower. 
 
\end{example}

}
\begin{table}
    \centering
    \hspace{-1.5 em}
	\begin{tabular}{ |c|c|c|c|c|c|c|c|c|c|c| } 
		\hline
		$\text{}$ & $D_1$ & $D_2$ & $D_3$ & $D_4$ & $D_5$ & $D_6$ & $D_7$ & $D_8$ & $D_9$ & $D_{10}$\\
		\hline
		$W_1$ & $1\times$ & $1\times$ & $1\checkmark$ & $1\checkmark$ & $\text{}$ & $\text{}$ & $\text{}$ & $\text{}$ &$\text{}$ & $\text{}$\\
		\hline 
		$W_2$ & $1\checkmark$ & $\text{}$ & $\text{}$ & $\text{}$ & $1\times$ & $1\times$ & $1\checkmark$ & $\text{}$ &$\text{}$ & $\text{}$\\
		\hline
		$W_3$ & $\text{}$ & $1\checkmark$ & $\text{}$ & $\text{}$ & $1\checkmark$ & $\text{}$ &$\text{}$ & $1\times$ & $1\times$ & $\text{}$\\
		\hline 
		$W_4$ & $\text{}$ & $\text{}$ & $1\times$ & $\text{}$ & $\text{}$ & $1\checkmark$  &$\text{}$ & $1\checkmark$ & $\text{}$ & $1\times$ \\
		\hline
		$W_5$ & $\text{}$ & $\text{}$ & $\text{}$ & $1\times$ & $\text{}$ & $\text{}$  & $1\times$ & $\text{}$ & $1\checkmark$ & $1\checkmark$\\
		\hline
	\end{tabular}
	\vspace{1 em}
	\caption{Assignment of data subsets (marked by $1\checkmark$ and $1\times$) to different workers in $(\frac{7}{10},3)$ feasible $(5,10,3,\frac{2}{5})$ GC scheme with data subsets marked by $1\checkmark$ having the corresponding gradients being directly transmitted by each corresponding worker. }
	\label{tab:example3}
\end{table}

\subsection{A construction based on $t$-designs}
\label{Sec:tdesigns}
In the earlier section, we have given a construction of combinatorial GC schemes which achieve the lower bound on the computation load. Now, we will give an alternate construction based on $t$-designs, which achieves the same computation load as that of combinatorial GC schemes and lower communication load. However, these schemes exist only on certain parameters, since $t$-designs themselves exist only for certain parameters.

\begin{definition}($t-(v,p,\lambda)$) \cite{10.5555-1202540}
A $t-(v,p,\lambda)$ design is an ordered pair $(S,B)$, where $S$ is a set of cardinality $v$, and $B$ is a family of $p$-subsets (called
blocks) of $S$ with the property that each $t$-subset of $S$ is
contained in precisely $\lambda$ blocks of $B$.
\end{definition}

Some of the well known properties of $t$-designs are listed below:\newline
    i) The parameters of the t-design are related as $\lambda=|B|\frac{{p \choose t}}{{v \choose t}}$.
    ii)  Every $t-(v,p,\lambda)$ design is also a $(t-1)-(v,p,\lambda^{*})$ with $\lambda^{*} = \lambda \times \frac{(v - t + 1)}{(p - t + 1)}$ as shown in \cite{10.5555-1202540}.
    iii) The complement of a $t-(v,p,\lambda)$ designs is a $t-(v,v-p, \lambda^*)$ design where $\lambda^* = \lambda \frac{{v-p \choose t}}{{p \choose t}}$. The complement design is obtained by considering family of $(v-p)$-subsets obtained by taking the complements of the $p$-subsets in the corresponding $t-(v,p,\lambda)$ design.
    

\begin{theorem}
\label{compub_load_tdesign}
If there exists a $t-(v,p,\lambda)$ design, then there exists an $(\alpha=1-\frac{{v-t \choose p}}{{v \choose p}},s=v-t)$-feasible $(n=v,k=|B|,m=\frac{|B|\times p}{v},l=\frac{p}{v})$ GC -scheme.
\end{theorem}

\begin{proof}

We propose an assignment scheme based on $t$-design, in which each element in $S$ corresponds to a distinct worker and each element in $B$ corresponds to a data-partition. Note that data-partition corresponding to each subset in $B$ is assigned to exactly those workers corresponding to the elements of the subset. Note that each worker transmits the gradients of all the data subsets computed by it. We show that the total number of data-partitions assigned to any set of $n-s$ workers matches with the lower bound on the computation load given in \eqref{final_lb}. Recall that the set of data-partitions assigned to worker $W_i$ is denoted by $\bigcup\limits_{j \in [m]} \text{supp}(A^i_j)$ which we denote by $E_j$ for the sake of brevity.
We will now compute the $|\bigcup\limits_{j \in N} E_j|$, where $N$ is an arbitrary subset of size $n-s = t$. Based on the property of the complementary design listed above, we have that $|\bigcap\limits_{j \in N} E_j^c| = \lambda^* = \lambda \frac{{v-p \choose t}}{{p \choose t}}$. Hence, we have that
\begin{eqnarray*}
|\bigcup\limits_{j \in N} E_j| & =& \left |\left (\bigcap\limits_{j \in N} E_j^c \right)^c \right | \\
& = & |B| - \lambda^*
 =  |B| - \lambda \frac{{v-p \choose t}}{{p \choose t}}
 =  |B| - |B| \frac{{p \choose t}}{{v \choose t}} \frac{{v-p \choose t}}{{p \choose t}} \\
& = & |B| \left ( 1 -  \frac{{v-p \choose t}}{{v \choose t}}\right )
 =  |B| \left ( 1 -  \frac{{v-t \choose p}}{{v \choose p}}\right ).
\end{eqnarray*}

Since each computed gradient is transmitted in this proposed scheme, we can say that the sum of $|B|.(1-\frac{{v-t \choose p}}{{v \choose p}})$ gradients can be computed whenever any set of $v-t$ workers straggle, thus it is $(\alpha=1-\frac{{v-t \choose p}}{{v \choose p}},s=v-t)$-feasible GC scheme. 
\end{proof}

\begin{example}

We will now give an example of the GC scheme described above based on $3-(8,4,1)$ design. It is also known as Hadamard $3-$ design. The set of blocks $B$ of the designs are given as follows:
\begin{eqnarray*}
&&\{\{1,2,5,6\} \{3,4,7,8\} \{1,3,5,7\} \{2,4,6,8\} \{1,4,5,8\} \{2,3,6,7\} \{1,2,3,4\} \\ && \{5,6,7,8\} \{1,2,7,8\} \{3,4,5,6\} 
 \{1,3,6,8\} \{2,4,5,7\} \{1,4,6,7\} \{2,3,5,8\}\}
\end{eqnarray*}

The assignment scheme for $n=8$, $s=5$ is as described in Table ~\ref{tab:example_t_design}. This is a scheme which achieves $\alpha=\frac{13}{14}$, $s=5$ with the computation cost of $\frac{1}{2}$ as proposed by the lower bound. In this case, the communication cost is 7, which is much lesser than the communication cost of the previous combinatorial GC scheme, i,e., ${{n-1 \choose y-1}} = 35$.

\begin{table}
    \centering
	\begin{tabular}{ |c|c|c|c|c|c|c|c|c|c|c|c|c|c|c| } 
		\hline
		$\text{Workers}$ & $D_1$ & $D_2$ & $D_3$ & $D_4$ & $D_5$ & $D_6$ & $D_7$ & $D_8$ & $D_9$ & $D_{10}$ & $D_{11}$ & $D_{12}$ & $D_{13}$ & $D_{14}$\\
		\hline
		$W_1$ & $\times$ & $\text{}$ & $\times$ & $\text{}$ & $\times$ & $\text{}$ & $\times$ & $\text{}$ &$\times$ & $\text{}$ &$\times$ & $\text{}$ & $\times$ & $\text{}$\\
		\hline 
		$W_2$ & $\times$ & $\text{}$ & $\text{}$ & $\times$ &$\text{}$ & $\times$ & $\times$ & $\text{}$ &$\times$ & $\text{}$ & $\text{}$ & $\times$ & $\text{}$ & $\times$\\
		\hline
		$W_3$ & $\text{}$ & $\times$ & $\times$ & $\text{}$ &$\text{}$ & $\times$ &$\times$ & $\text{}$ & $\text{}$ & $\times$ & $\times$ & $\text{}$ & $\text{}$ & $\times$\\
		\hline 
		$W_4$ & $\text{}$ & $\times$ & $\text{}$ & $\times$ & $\times$ & $\text{}$  &$\times$ & $\text{}$ & $\text{}$ & $\times$ & $\text{}$ & $\times$ & $\times$ & $\text{}$\\
		\hline
		$W_5$ & $\times$ & $\text{}$ & $\times$ & $\text{}$ & $\times$ & $\text{}$  & $\text{}$ & $\times$ & $\text{}$ & $\times$ & $\text{}$ & $\times$ & $\text{}$ & $\times$\\
		\hline
		$W_6$ & $\times$ & $\text{}$ & $\text{}$ & $\times$ & $\text{}$ & $\times$ & $\text{}$ & $\times$ &$\text{}$ & $\times$ & $\times$ & $\text{}$ & $\times$ & $\text{}$ \\
		\hline 
		$W_7$ & $\text{}$ & $\times$ & $\times$ & $\text{}$ &$\text{}$ & $\times$ & $\text{}$ & $\times$ &$\times$ & $\text{}$ & $\text{}$ & $\times$ & $\times$ & $\text{}$\\
		\hline
		$W_8$ & $\text{}$ & $\times$ & $\text{}$ & $\times$ & $\times$ & $\text{}$ &$\text{}$ & $\times$ & $\times$ & $\text{}$ & $\times$ & $\text{}$ & $\text{}$ & $\times$\\
		\hline 
	\end{tabular}
	\vspace{1 em}
	\caption{Assignment of data subsets to different workers for $(\frac{13}{14},5)$- feasible $(8,14,7,\frac{1}{2})$ GC scheme}
	\label{tab:example_t_design}
\end{table}

\end{example}

%% file: cyclic_scheme.tex
\section{Cyclic $(\alpha,s)$-feasible GC schemes}{\label{cyclic_scheme_section}}

In the previous section, we presented two schemes which achieves minimum computation load at the cost of high communication load and large number of data partitions. In this section, we will consider $(\alpha,s)$-feasible GC schemes, when the number of data subsets is restricted to $n$ (the number of workers) and the assignment of data subsets is cyclic. We are interested in the cyclic assignment based GC schemes because they have been shown to be optimal for the case of gradient coding with full recovery \cite{tandon2017gradient}, \cite{raviv2020gradient}. Also, for the case of gradient coding with partial recovery, {random cyclic shift based schemes have been proposed in \cite{9641837}}, though their optimality has not been shown. We provide two schemes based on cyclic assignment of data subsets of workers. The first scheme requires that the parameters of the GC scheme satisfy a certain divisibility criterion, in which case we show that there exists an  $(\alpha,s)$-feasible GC scheme with a communication load of 1. We then show that whenever the divisibility criterion is not met, cyclic schemes cannot achieve the desired computation load, when the communication load is 1. Finally, we show that there exists an $(\alpha,s)$-feasible cyclic GC scheme with a communication load of 2 {for some subset of parameter values}. 

\begin{definition}{\label{cyclic_scheme_defn}}
(Cyclic GC scheme): We define a $(n,n,m,l)$ GC scheme as a cyclic GC scheme if worker $W_1$ is assigned the data subsets from $D_1$ to $D_{l\times n}$, worker $W_2$ is assigned the data subsets from $D_2$ to $D_{l\times n+1}$ and in general worker $W_i$ is assigned the data subsets from $D_i$ to $D_{1+((l\times n+i-2) \mod \ n)}$.

\remove{
the following conditions are satisfied for some integer $r \in [n]$.

\begin{itemize}
    \item  $A^i_{j,t}=0$ if $t> (1+((i+r-2)\mod \ n))$ or $t <i $, $\forall i \in [n] \forall j \in [m]$
    \item  $\forall t = (1+((i+u-2)\mod \ n)) \ \ \\exists j \in [m]$ satisfying $A^i_{j,t}\neq 0$, $\forall i \in [n] \forall u \in [r]$. 
\end{itemize}
}
\end{definition}

\remove{
Note that the definition above ensures that worker $W_i$ is assigned the data subsets from $D_i$ to $D_{1+((i+r-2)\mod \ n)}$ (in cyclic ordering) and the gradients of each of these data subsets assigned to a worker is present in one or more linear combinations transmitted by the worker.
}

\begin{theorem}{\label{cyclic_constr_comm_1}}
 There exists an $(\alpha,s)$-feasible $(n,n,1,\frac{s+1+\beta-n}{n})$ cyclic GC scheme with $\beta =\ceil{ \alpha.n}$ for every $n,s,\alpha$ if $s+1+\beta-n$ divides $\beta$.  
\end{theorem}

\begin{proof}
{We follow the assignment scheme as described in Definition~\ref{cyclic_scheme_defn} and each worker is assigned exactly $s+1+\beta-n$ data subsets. Each worker transmits the sum of the gradients of all the data subsets assigned to it.} To show that the scheme is $(\alpha,s)$-feasible, we show that we can recover the sum of $\beta = \lceil \alpha .n \rceil$ data subsets in the presence of any $s$ stragglers. Based on the straggler pattern, we pick a subset of $n-s$ non-straggling workers of size $\frac{\beta}{r}$, such that $r$ data subsets assigned to these workers are mutually disjoint and give an algorithm to identify these workers in Appendix \ref{cyclic_constr_comm_1_proof}.
\end{proof}
The following theorem shows that if $s+1+\beta-n$ does not divide $\beta$, no $(\alpha,s)$ feasible $(n,n,1,\frac{s+1+\beta-n}{n})$ cyclic GC  scheme exists. 

{
\begin{theorem}{\label{cyclic_scheme_lower_bound}}
  There exists no $(\alpha,s)$-feasible $(n,n,1,\frac{t}{n})$ cyclic GC scheme if $s+1+\beta-n$ does not divide $\beta$ and $t \leq s+1+\beta-n$ where $\beta=\ceil{\alpha.n}$ and $\beta \leq n-1$.
\end{theorem}
}
\begin{proof}
Suppose there exists an $(\alpha,s)$ feasible $(n,n,1,\frac{s+1+\beta-n}{n})$ cyclic GC scheme, thus each worker has access to exactly $v= s+1+\beta-n$ data subsets. Consider any 2 set of consecutive data subsets $D_i$ and $D_{1+(i\mod \ n)}$. Choose a set of $s-1$ consecutive workers from $W_{1+((i-s)\mod \ n)}$ to $W_{i-1}$ and another worker $W_{i+1}$ and straggle them. Since the master should be able to compute a sum of atleast $\beta$ gradients from the results received from each worker except the set of $s$ workers defined above, the coefficient of the gradients of data subsets $D_i$ and $D_{1+(i\mod \ n)}$ transmitted by worker $W_i$ have to be the same. This is because the master has access to exactly $\beta+1$ gradients and the gradient of data subsets $D_i$ and $D_{1+((i)\mod \ n)}$ is computed only by $W_i$ amongst the set of non-straggling workers.

Using a very similar line of argument, we can show that the coefficient of the gradients of data subsets corresponding to $D_i$ and $D_{1+(i \mod \ n)}$ transmitted by any other worker which has access to both of them must also be the same. This can be argued for every $i \in [n]$. This would imply that each worker just transmits the sum of all the gradients assigned to it as per the cyclic GC scheme discussed above.

Now suppose the set of non-straggling workers is denoted by $W_1,W_2,..W_{n-s}$. Clearly under these set of workers the master would have access to exactly gradients of $\beta$ data subsets. Suppose we denote the first row of the matrix $A_i$ for $i=1,2...,n-s$ as $v_i$. Since the master node should be able to compute the sum of the gradients of first $\beta$ data-sets from transmissions by workers $W_1,W_2,..W_{n-s}$, $v=[\underbrace{1,1,..1}_{\beta},\underbrace{0,0,..0}_{n-\beta}]$ must lie in the span of $\{v_i\}$. Also note that vector $v_i$ has consecutive ones from position $i$ to $(1+((i+r-2)\mod \ n))$ for $r=s+1+\beta-n$ rest all zeroes.

Suppose $v=\sum_i c_i v_i$ for some $c_i \in \mathbb{R}$. This would imply that $c_1=1$,$c_2=0$,...,$c_{s+1+\beta-n}=0$
,$c_{s+2+\beta-n}=0$,..$c_{2s+2+\beta-n}=0$ and so on. More generally, $c_{i}=1$ if $i \mod \ r=1$ else 0 where $r=s+1+\beta-n$. Now we can substitute the $\{c_i\}$ in the equation $v=\sum_i c_i$ and observe that it can't be satisfied if $s+1+\beta-n$ does not divide $\beta$. Thus the master cannot recover the sum of $\beta$ data subsets and hence such a cyclic GC scheme is not $(\alpha,s)$ feasible

\remove{

However, since $s+1+\beta-n$ does not divide $\beta$, the master cannot recover the sum of gradients of $\beta$ data subsets from the sum of $s+1+\beta-n$ gradients transmitted by each worker.

}
Now we consider the case of $t < s+1+\beta-n$. For a $(\alpha,s)$-feasible $(n,n,1,\frac{t}{n})$ cyclic GC scheme, the set of data subsets assigned to any of $W_1,W_2,\ldots,W_{n-s}$ is given by $D_1,D_2,\ldots, \\D_{t+n-s-1}$ whose cardinality is clearly smaller than $\beta$. Thus we cannot compute the sum of gradients of any set of $\beta$ data-subsets when $W_1,W_2,...,W_{n-s}$ do not straggle, hence no $(\alpha,s)$-feasible $(n,n,1,\frac{t}{n})$ cyclic GC scheme exists for $t \leq s+1+\beta-n$.  \end{proof}

However, we can show that a cyclic $(\alpha,s)$ feasible cyclic $(n,n,1,\frac{s+1+\beta-n}{n})$ GC scheme is always possible under a communication cost of 2  {for certain parameters}.
\begin{theorem}\label{cyclic_constr_comm_2}
 There exists an $(\alpha,s)$-feasible $(n,n,2,\frac{s+1+\beta-n}{n})$ cyclic GC scheme with $\beta =\ceil{ \alpha.n}$ for every $n,s,\alpha$ satisfying {$r - (\beta \mod r) \leq n - \beta$ where $r$= $s+1+\beta-n$}. 
\end{theorem}

\begin{proof}
 If $s+1+\beta-n$ divides $\beta$, then the scheme described in proof of Theorem \ref{cyclic_constr_comm_1} achieves a communication load of 1. Else consider the following scheme described below.

{We follow the assignment scheme as described in Definition~\ref{cyclic_scheme_defn} and each worker is assigned exactly $s+1+\beta-n$ data subsets. Each worker transmits the sum of the gradients of all the data subsets assigned to it and the sum of first $x$ data subsets assigned to it where $x$ denotes the remainder when $\beta$ is divided by $s+1+\beta-n$. For example worker $W_1$ transmits the sum of the gradients of the data subsets from $D_1$ to $D_{s+1+\beta-n}$ and the sum of the gradients of the data subsets from $D_1$ to $D_x$. The correctness of the construction scheme described above is given in Appendix ~\ref{cyclic_constr_comm_2_proof}} \end{proof}

%% file: scheme_low_comp_comm_1.tex
\section{$(\alpha,s)$-feasible $(n,n,1,l)$ GC schemes under low computation load}
\label{sec:lowcomputation}
In this section, we consider the problem of partial gradient recovery under the restriction of $k=n$ data subsets, and focus on the regime with communication load $m = 1$ and a small computation load $l$. We begin with a simple lemma about the case of $l=1/n$ which is the minimum possible computation load.
\begin{lemma}
For any $(\alpha,s)$-feasible $(n,n,1,\frac{1}{n})$ GC scheme, we have $s \le n - \beta$ for $\beta = \ceil{\alpha.n}$. Furthermore, there exists a simple $(\alpha, s = n - \beta)$-feasible $(n,n,1,\frac{1}{n})$ GC scheme.
\end{lemma}
\begin{proof}
For $l = 1/n$, each worker is assigned at most one data subset and thus when there are $s$ stragglers, the master node can hope to recover the sum of the gradients corresponding to at most $n-s$ data subsets. Then from the definition of an $(\alpha, s)$-feasible GC scheme, we have $\alpha. n \le n - s$ which in turn implies $s \le n - \ceil{\alpha.n}$. Finally, the trivial scheme which assigns a unique data subset to each worker node and each non-straggler node simply computes and transmits the corresponding gradient to the master node is indeed $(\alpha,n - \ceil{\alpha.n})$-feasible.
\end{proof}
The next two results consider the impact of allowing for more stragglers on the computation load $l$. 
\begin{theorem}{\label{lower_bound1_low_comp}}
For $\beta = \ceil{\alpha.n}$, consider any $(\alpha,s=n-\beta + 1)$ feasible $(n,n,1,l)$ GC. Then the following hold true.
 \begin{itemize}
 \item If $\beta$ is even and $\beta \leq n-1$, then $l\geq \frac{s+1+\beta-n}{n}=\frac{2}{n}$. Furthermore, there exists a cyclic scheme which achieves $l = \frac{2}{n}$.
     \item If $\beta$ is odd and $\beta \leq n-1$, then $l>\frac{s+1+\beta-n}{n}=\frac{2}{n}$.
 \end{itemize}
\end{theorem}

\begin{proof}
The first half  of the first statement follows from Theorem \ref{final_lb} by showing that {inequality \eqref{lb_equation}} is unsatisfied when $y=1$. The existence of a cyclic scheme for even $\beta$ with $l = 2/n$ can be shown using Theorem~\ref{cyclic_constr_comm_1}. Finally, the proof of the second part of the theorem can be found in Appendix~\ref{lower_bound1_low_comp_section}.  
\end{proof}

{For $\beta=\ceil{\alpha.n}$, Theorem~\ref{lower_bound1_low_comp} shows that no $(\alpha,s=n-\beta+1)$ feasible $(n,n,m,\frac{2}{n})$ GC exists for communication cost $m=1$, while Theorem~\ref{cyclic_constr_comm_2} shows the existence of $(\alpha,s=n-\beta+1)$ feasible $(n,n,m,\frac{2}{n})$ GC for communication cost $m=2$, thus implying a tradeoff between communication cost and computation load.}
\remove{
For even $\beta$ and $s=1-\beta+n$, we can show that the above inequality would reduce to $l> \frac{1}{n}$ as giving $1$ data-part to every worker won't work since straggling of any $s$ workers would only the availability of the gradients of $n-s=\beta-1$ data-parts at the master node. The cyclic coding scheme discussed in previous section achieves a computation load of $\frac{2}{n}$ for $s=1-\beta+n$ and even $\beta$.
}
\remove{

The above theorem states that there must exist a worker with 3 data-parts assigned to it if $s+1+\beta-n=2$ and $\beta$ is odd. The following theorem gives a construction to achieve the above mentioned lower bound. 
}
\begin{theorem}{\label{lower_bound2_low_comp}}
For $\beta = \ceil{\alpha.n}$ and $\beta \leq n-1$, consider any $(\alpha,s>n-\beta+1)$ feasible $(n,n,1,l)$ GC. Then $l>\frac{(n-\beta+1)+1+\beta-n}{n}=\frac{2}{n}$.
\end{theorem}

The proof of this result can be found in Appendix~\ref{lower_bound2_low_comp_section}.

\begin{table}
    \centering
\begin{tabular}{|c|c|c|c|} 
 \hline
 Parameters& Scheme & Comm. Cost & Comp. Load \\
 \hline
 $n=7$, $\alpha=6/7$, $s=3$ & Naive GC & $1$ & $24/49$ \\
 \hline
 & Cyclic GC & $1$ & $3/7$ \\ 
 \hline
 & Combinatorial GC & 6 & $2/7$ \\
 \hline
 \hline
 $n=9$, $\alpha=7/9$, $s=4$ & Naive GC & $1$ & $35/81$ \\
 \hline
  & Cyclic GC & $2$ & $3/9$ \\ 
\hline
 & Combinatorial GC & 8 & $2/9$ \\
\hline
\end{tabular}
\vspace{1 em}
   \caption{Comparison of communication cost and computation load incurred by Uncoded, Cyclic GC and Combinatorial GC schemes for two example parameters.}
    \label{comp_cyclic_comb}
\end{table}
\noindent{A comparison of the computation load and communication cost incurred by naive GC scheme, combinatorial GC scheme and cyclic GC scheme for some parameters is given in Table \ref{comp_cyclic_comb}. We see that combinatorial GC scheme has the least computation load. Cyclic GC schemes have lower computation load than naive GC scheme with low communication cost. }

\section{$(\alpha,s)$ feasible GC schemes for Intermediate Points}{\label{compload_inter_points}}

In the previous sections, we have seen broadly two classes of GC schemes. One class of schemes is based on large number of data-partitions and the other is based on cyclic assignment of data-partitions. We will refer to the first class as combinatorial GC schemes and the second one as cyclic GC schemes. The combinatorial GC schemes have low computation load and high communication cost, whereas the cyclic GC schemes have high computation load and low communication cost. In this section, we will give a gamut of $(\alpha,s)$ feasible GC schemes, which achieve different intermediate operating points of the computation and communication load.

\begin{theorem}{\label{high_comm_cyclic_middle}}
There exists an $(\alpha,s)$ feasible $(n,\frac{n}{t}{{n-\delta+y-1 \choose y-1}},\frac{\delta}{t}{{n-1-\delta+y \choose y-1}},\frac{\delta}{n})$ GC scheme for all positive integers $\delta,y$ such that $\delta\leq s$ and $t=\text{gcd}(\delta,y)$ satisfying the following inequality:
 \begin{equation}{\label{high_comm_cyclic_middle_eqn1}}
  \frac{y{s-\delta+y \choose y}}{{n{n-\delta+y-1 \choose y-1}}}\leq (1-\alpha).   
 \end{equation}
\end{theorem}

\remove{
except with a bit higher communication cost of $\delta$ (transmitting gradients of all the data-partitions assigned). We denote $\delta^{*}(y)$ as the minimum value of $\delta$ satisfying the inequality \eqref{high_comm_cyclic_middle_eqn1} for a given value of $y$. We can show that the computation load $\frac{\delta^{*}(y)}{n}$ decreases from the cyclic computation load in Sec ~\ref{cyclic_scheme_section} to Sec ~\ref{compload_ub_section} to as $y$ increases from 1 (stated as a corollary below) and the communication cost (as a function of $\delta^{*}$ and $y$) decreases as $y$ increases from 1.
}

\begin{proof}
The proof of the theorem is described in four steps below:

\noindent {\bf Step 1- Describing data-partitions and assignment for $t=1$}: Recall that for the scheme in Section ~\ref{compload_ub_section}, each data partition was denoted by a set of $y$ indices. In this scheme, we denote each data-partition by a unique list of $y$ elements (denoted by $[c_1,c_2,\ldots,c_y]$). Let us aprioiri choose integers $\{\gamma_i\}_{i=1}^{y}$ such that $\sum_i{\gamma_i}=\delta$. We impose the additional constraint that $(c_{1+(i \mod n)}-c_i \mod n) \geq \gamma_i$ $\forall i \in [y]$. Note that corresponding to each such unique list, we have a unique data-partition. Now a data-partition denoted by the list ($[c_1,c_2,\ldots,c_y]$) is assigned to worker $W_j$ iff $0 \leq ((j - c_i) \mod n) \leq \gamma_i-1$ for some $i \in [y]$. We individually transmit the gradients of all the data-partitions assigned to a worker. We now argue that such a scheme is $(1-\frac{y{s-\delta+y \choose y}}{{n.{n-\delta+y-1 \choose y-1}}},s)$ feasible and hence, such a $(\alpha,s)$ GC-scheme exists whenever $\frac{y{s-\delta+y \choose y}}{{n.{n-\delta+y-1 \choose y-1}}}\leq (1-\alpha)$.

\noindent {\bf Step 2- Computing the total number of data-partitions for $t=1$}: First we compute all possible lists with the first element 1. Thus, we have $c_2 \geq \gamma_1+1$ and $c_y \leq n-\gamma_y+1$ with $c_{i+1}-c_i \geq \gamma_i$. By a combinatorial argument, we show that the number of such lists is given by ${{n-\gamma_y-(\sum_{i=1}^{y-1}\gamma_i)+(y-1) \choose y-1}}= {{n-\delta+y-1 \choose y-1}}$ as $\delta= \sum_{i=1}^{y}\gamma_i$. Since any of the $n$ indices can take the first position in list, the total number of such lists is $n.{{n-\delta+y-1 \choose y-1}}$. Hence, the total number of data-partitions is $n{{n-\delta+y-1 \choose y-1}}$.

\noindent {\bf Step 3- Assignment and Merging for general $t$}: Since $y$ and $\delta$ both are multiples of $t$, denote integers $\tilde{y}=\frac{y}{t}$ and $\tilde{\delta}=\frac{\delta}{t}$. In this case we choose $\beta_1,\beta_2,\ldots,\beta_{\tilde{y}}$ s.t. $\sum_{i=1}^{y/t} \beta_i= \delta/t$. Now $\lambda_i = \beta_{1+(i-1 \mod \tilde{y})}$ in which case $\sum_i {\lambda_i}=\delta$. Now consider any data-partition denoted by the list $[c_1,c_2,\ldots,c_y]$ which is assigned to workers in set $\tilde{W}$. We now argue that the data-partition denoted by the list $[c_{\tilde{y}+1},c_{\tilde{y}+2},\ldots,c_y,\ldots,c_{\tilde{y}}]$ would also be assigned exactly to workers in $\tilde{W}$. This follows since $\lambda_{(1+(i+\tilde{y}-1 \mod y))}=\lambda_i$ and the entries in the second list are shifted by $\tilde{y}$ compared to the first. In general, every data-partition denoted by the list $[c_{i.(\tilde{y})+1},c_{i.\tilde{y}+2},\ldots,c_y,\ldots,c_{i.\tilde{y}}]$ is also assigned exactly workers in $\tilde{W}$ for every $i \in [t]$. Now we merge all such data-partitions which are assigned to the same set of workers. We now argue that after merging all such data-partitions, each worker would be assigned the same number of data-partitions.

\noindent {\bf Step 4- Lower bound on the number of data-partitions assigned to $n-s$ non-straggling workers}: The number of data-partitions which are assigned to any $n-s$ non-straggling workers is lower bounded by  $n{{n-\delta+y-1 \choose y-1}}-y{{s-\delta+y \choose y}}$ and the proof of this part is given in Appendix \ref{high_comm_cyclic_appendix}.

\end{proof}

\begin{example}
We will now consider a $(\alpha=13/15, s=3)$ GC scheme with parameters $n=5$, $s=3$, $y=2$ and $\delta=3$ as described in Theorem ~\ref{high_comm_cyclic_middle}. For the case of $\delta=3$, we choose $\gamma_1=1$ and $\gamma_2=2$. The set of all lists denoting data-partitions can be written as\\ $\{[1,2],[1,3],[1,4],[2,3],[2,4],[2,5],[3,4],[3,5],[3,1],[4,5],[4,1],[4,2],[5,1],[5,2],[5,3]\}$ which are denoted (in that order) by $\{1,2,3,\ldots,15\}$ in the following table. Each worker transmits all the gradients of the data-partitions assigned to it.

\begin{table}[!h]
    \centering
	\begin{tabular}{|c|c|c|c|c|c|c|c|c|c|c|c|c|c|c|c|c| } 
		\hline
		$\text{Workers}$ & $D_1$ & $D_2$ & $D_3$ & $D_4$ & $D_5$ & $D_6$ & $D_7$ & $D_8$ & $D_9$ & $D_{10}$ & $D_{11}$ & $D_{12}$ & $D_{13}$ & $D_{14}$ & $D_{15}$\\
		\hline
		$W_1$ & 1 & 1 & 1 & $\text{}$ & $\text{}$ & 1 & $\text{}$ & 1 & 1 & 1 & 1 & $\text{}$ & 1 & $\text{}$ & $\text{}$ \\
		\hline 
		$W_2$ & 1 & $\text{}$ & $\text{}$ & 1 & 1 & 1 & $\text{}$ & $\text{}$ & 1 & $\text{}$ & 1 & 1 & 1 & 1 & $\text{}$\\
		\hline
		$W_3$ & 1 & 1 & $\text{}$ & 1 & $\text{}$ & $\text{}$ & 1 & 1 & 1 & $\text{}$ & $\text{}$ & 1 & $\text{}$ & 1 & 1 \\
		\hline 
		$W_4$ & $\text{}$ & 1 & 1 & 1 & 1 & $\text{}$ & 1 & $\text{}$ & $\text{}$ & 1 & 1 & 1 & $\text{}$ & $\text{}$ & 1\\
		\hline
		$W_5$ & $\text{}$ & $\text{}$ & 1 & $\text{}$ & 1  & 1 & 1 & 1 & $\text{}$ & 1 & $\text{}$ & $\text{}$ & 1 & 1 & 1 \\
	
		\hline
	\end{tabular}
	\vspace{1 em}
	\caption{Assignment of data subsets to different workers in $(\frac{13}{15},3)$ feasible $(5,15,9,\frac{3}{5})$ GC scheme as described in Theorem ~\ref{high_comm_cyclic_middle}.}
	\label{tab:example4}
\end{table}

\end{example}

\begin{remark}  We denote $\delta^{*}(y)$ as the minimum value of $\delta$ satisfying the inequality \eqref{high_comm_cyclic_middle_eqn1} for a given value of $y$.
Note that for the case of $\delta=y$, the computation load of this scheme is same as that of the scheme described in Section ~\ref{compload_ub_section} where as for the case of $y=1$ it reduces to that of the cyclic schemes as in Section ~\ref{cyclic_scheme_section}.  We will show that the computation load $\frac{\delta^{*}(y)}{n}$ decreases from the cyclic computation load in Sec ~\ref{cyclic_scheme_section} to the optimal computation load as $y$ increases from 1 to $y^*$ where $y^*$ is such that $\delta^*(y^*) = y^*$. Thus, these give a set of gradient codes with computation load between the cyclic GC schemes and combinatorial GC schemes. Also note that the communication cost of these schemes typically lies between the two extremes, but may not be monotonic with the computation load.
\end{remark}

\begin{lemma}
Suppose, we denote $\delta^{*}(y)$ as the smallest $\delta$ satisfying \eqref{high_comm_cyclic_middle_eqn1} for a given $y$, then the computation load $\frac{\delta^{*}(y)}{n}$ decreases with $y$.
\end{lemma}

\begin{proof}
Suppose we denote $\delta^{*}(y_1)y_1$ as $r_1$. Similarly, we denote $\delta^{*}(y_1+1)(y_1+1)$ as $r_2$. Note that the inequality for the case $y=y_1$ is $\frac{y_1{s-r+y \choose y_1}}{{n{n-r+y_1-1 \choose y_1-1}}}\leq (1-\alpha)$ and in the second case is $\frac{(y_1+1){s-r+y_1+1 \choose y_1+1}}{{n{n-r+y_1 \choose y_1}}}\leq (1-\alpha)$. However, on dividing the LHS of the first inequality with the second, we obtain $\Bigl(\frac{y_1}{y_1+1}\Bigr)\Bigl(\frac{y_1+1}{s-r+y_1+1}\Bigr)\Bigl(\frac{n-r+y_1}{y_1}\Bigr)=\frac{n-r+y_1}{s-r+y_1+1}>1$, thus the smallest value of $r$ satisfying the first inequality must be larger than or equal to the second i.e $r_1 \geq r_2$ as $r_1$ must be a solution to the second inequality too. \end{proof}

\begin{figure}[ht]
\begin{minipage}[b]{0.45\linewidth}
\centering
\includegraphics[scale =0.5]{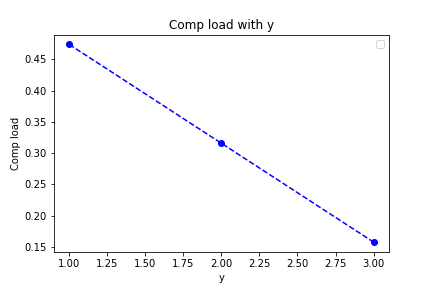}
\caption{Computation load with $y$ for $n=19$, $s=10$, $\alpha=0.87$}
\label{comp_load_n_19_s_10_alpha_87}
\end{minipage}
\hspace{0.5cm}
\begin{minipage}[b]{0.45\linewidth}
\centering
\includegraphics[scale = 0.5]{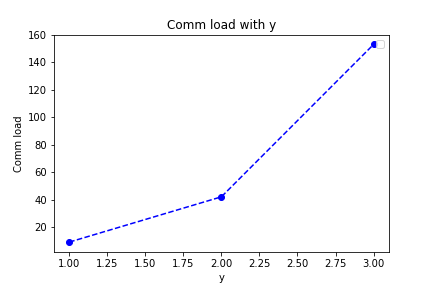}
\caption{Communication cost with $y$ for $n=19$, $s=10$, $\alpha=0.87$}
\label{comm_cost_n_19_s_10_alpha_87}
\end{minipage}
\end{figure}

We increase $y$ from 1 in which case the assignment scheme is same as that described in Sec~\ref{cyclic_constr_comm_1}. We compute the communication cost and computation load according to Theorem~\ref{high_comm_cyclic_middle} corresponding to the smallest $\delta$ satisfying \eqref{high_comm_cyclic_middle_eqn1} ($\delta^{*}(y)$) for given $y$. We increase $y$ until $\delta^{*}(y)$ reduces to $y$ in which case the assignment scheme is same as the combinatorial GC scheme described in Sec ~\ref{compload_ub_section}. In Fig.~\ref{comp_load_n_19_s_10_alpha_87} and Fig.~ \ref{comm_cost_n_19_s_10_alpha_87}, we plot the communication and computation cost according to Theorem~\ref{high_comm_cyclic_middle} for the case of $n=19, s=10\text{ and }\alpha=0.87$. Note that the values of $\delta^{*}(y)$ takes values $[9,6,3]$ for $y = [1,2,3]$.

\section{Stochastic models for computation time}
\label{sec:stochastic}
In the sections above, we have analyzed the computation load of various schemes in terms of the number of partial gradients that a worker has to compute, for a given accuracy $\alpha$ and straggler tolerance $s$. In this section, we will instead take an average computation delay viewpoint, wherein we consider a stochastic model for the computation time of a gradient at a server and then consider the total expected computation delay of a scheme, for a given accuracy $\alpha$, as a metric of performance. 

There are several popular stochastic models for server computation time \cite{9174030}. Let the computation time of one partial gradient on a server be a random variable $X$; we consider the following distribution for $X$: 

\textit{Pareto, $X \sim \text{Pareto}(\lambda, \rho)$} : Support of $X$ is $[\lambda, \infty)$ where $\lambda$ is the minimum computation time. The tail
distribution is given as $\text{Pr}(X > x)$ = $(\frac{\lambda}{x})^{\rho}$ for $x > \lambda$, which models a heavy tail and $\rho$ is known as the tail index. A smaller $\rho$ corresponds to a heavier tail, which then implies a higher probability of straggling.

\textit{(Shifted)- Exponential, $X \sim \text{S-Exp}(\Gamma, W)$} : The support of $X$ is $[\Gamma, \infty)$ where $\Gamma$ is the minimum service time. The tail
distribution is given as $\text{Pr}(X > x)$ = $e^{-\frac{x-\Gamma}{W}}$ for $x> \Gamma$. Note that the larger the $W$ is, the higher the probability of `straggling', i.e., a large computation time.

We will now describe a model for the total delay incurred for computing multiple partial gradients at a server \cite{9174030}, which essentially describes how the delay scales with the computation load at the server. Delays across servers are assumed to be independent. 

\textit{Data-Dependent Scaling}: Here the assumption is that each gradient computation takes a fixed deterministic time $\Delta$ and in addition, there is some random delay $X$ which is independent of the total number of gradients $l$. Thus, the total time needed to complete the computation of $l$ partial gradients at a server is given by $Y = l\times \Delta + X$.

\textit{Server-dependent Scaling}: Here the assumption is that each gradient computation takes a fixed random time $X$ irrespective of the total number of gradients $l$. Thus, the total time needed to complete the computation of $l$ partial gradients at a server is given by $Y = l\times X$.

For a given $\alpha$, different $(\alpha, s)$-feasible schemes are now compared in terms of the expected total computation time. To illustrate this, we will consider the following two schemes. 
\begin{enumerate}
    \item \textit{Scheme $1$}: A benchmark policy is the uncoded scheme where each worker computes gradients corresponding to a single data partition, with the total number of data partitions $k$ being equal to $n$, and the master just waits for the fastest $\alpha n$ workers to complete their computations. Thus, this policy corresponds to an $(\alpha, s_1 = n(1-\alpha))$-feasible $(n, n, 1, \frac{1}{n})$ GC scheme. 
    \item \textit{Scheme $2$}: As a representative of the various redundancy and coding-based schemes proposed in the previous sections, we consider the scheme with optimal computation load in Theorem~\ref{compload_ub} with $y=2$, so that the total number of data partitions $k$ is $n(n-1)/2$ and each data partition is assigned to two workers. From Theorem~\ref{compload_ub}, it suffices for the master node to receive the computation results from any $n - s_2$ workers, where $s_2$ satisfies $\frac{(s_2(s_2-1))}{n(n-1)}=(1-\alpha)$. Thus, this policy corresponds to an $(\alpha, s_2)$-feasible $(n,\frac{n.(n-1)}{2},n-1,\frac{2}{n})$ GC scheme. Note that when $n$ is large, we have $s_2 \approx n\sqrt{(1-\alpha)}$.  
\end{enumerate}
Assuming the complete dataset to contain $d$ points, the two schemes mentioned above differ in the number of gradients each worker has to complete and also the number of workers the master node waits for. Next, using expressions derived in \cite{9174030}, we compare the expected overall computation time of the two schemes under the delay model discussed before. Also let $H_n$ denote the harmonic sum from 1 to $n$ i.e. $H_n = \sum_{i=1}^{n} \frac{1}{i}$. Note that $H_n \approx \log n$ for $n$ large.

\subsection{Pareto distribution under the data-dependent scaling case} 
Under this model, the computation time at a worker for evaluating $l$ gradients is given by $Y = l.\Delta + X$, where $X \sim \text{Pareto}(\lambda, \rho)$, so that {$Y\sim l\Delta + \text{Pareto}(\lambda, \rho)$} . For Schemes $1$ and $2$, the corresponding expected overall computation delays are given by $\frac{d}{n}\Delta + \lambda.\frac{n\,!}{(s_1)\,!}\frac{\Gamma(s_1+1-\frac{1}{\rho})}{\Gamma(n+1-\frac{1}{\rho})} \approx \frac{d}{n}\Delta+\lambda (\frac{n}{s_1})^{\frac{1}{\rho}}$ and $\frac{2d}{n}\Delta + \lambda.\frac{n\,!}{(s_2)\,!}\frac{\Gamma(s_2+1-\frac{1}{\rho})}{\Gamma(n+1-\frac{1}{\rho})} \approx \frac{2d}{n}\Delta+\lambda (\frac{n}{s_2})^{\frac{1}{\rho}}$ respectively. Thus, we can see that the delay under Scheme $1$ is larger than that for Scheme $2$ when $\lambda. (1-\alpha)^{-\frac{1}{2.\rho}}\left((1-\alpha)^{-\frac{1}{2\rho}} - 1\right) > \frac{d}{n}\Delta$. This condition is satisfied when $\alpha$ is close to $1$, $\rho$ is small, or $\Delta$ is small. 

\subsection{Shifted- Exponential distribution under the data-dependent scaling case}

Under this model, the computation time at a worker for evaluating $l$ gradients is given by $Y = l.\Delta + X$, where $X\sim \text{S-Exp}(\Gamma,W)$, so that $Y\sim \text{S-Exp}(l.\Delta+\Gamma, W)$. For Schemes $1$ and $2$, the corresponding expected overall computation delays are given by ${\frac{d}{n}.\Delta+\Gamma+W(H_{n}-H_{s_1}) \approx \frac{d}{n}.\Delta + \Gamma+W\log(n/s_1)}$ and $\frac{2d}{n}.\Delta+\Gamma+W(H_{n}-H_{s_2}) \approx \frac{2d}{n}.\Delta + \Gamma+W\log(n/s_2)$ respectively. Thus, we can see that the delay under Scheme $1$ is larger than that for Scheme $2$ when $W\log(s_2/s_1) > \frac{d}{n}.\Delta$. Using $s_1\sim n {(1-\alpha)}$ and $s_2\sim n \sqrt{(1-\alpha)}$, a sufficient condition for our proposed scheme to have a lower expected computation delay than the benchmark uncoded scheme is  $\frac{W}{2}\log(\frac{1}{1-\alpha})>\frac{d}{n}\Delta$. This condition is satisfied, for example, when the desired accuracy $\alpha$ is high or when the deterministic part of the overall computation delay $\frac{d}{n}\Delta$ is small. 

\subsection{Pareto distribution under server dependent scaling model}
Under this model, the computation time at a worker for evaluating $l$ gradients is given by $Y = l.X$ where $X \sim \text{Pareto}(\lambda, \rho)$, so that $Y \sim \text{Pareto}(l.\lambda, \alpha)$. For Schemes $1$ and $2$, the corresponding expected overall computation delays are given by {$\frac{d}{n}.\lambda.\frac{n\,!}{s_1\,!}\frac{\Gamma(s_1+1-\frac{1}{\rho})}{\Gamma(n+1-\frac{1}{\rho})} \approx \frac{d}{n}\lambda (\frac{n}{s_1})^{\frac{1}{\rho}}$ and $\frac{2d}{n}.\lambda.\frac{n\,!}{s_2\,!}\frac{\Gamma(s_2+1-\frac{1}{\rho})}{\Gamma(n+1-\frac{1}{\rho})} \approx \frac{2d}{n}\lambda (\frac{n}{s_2})^{\frac{1}{\rho}}$} respectively. Thus, for {$(1-\alpha)^{-\frac{2}{\rho}}> 2$}, the overall computation delay under the uncoded policy Scheme $1$ is larger than that under the proposed policy Scheme $2$. This condition is satisfied when when the desired accuracy $\alpha$ is high or when the tail index $\rho$ is small, which corresponds to a heavier tail for the distribution of the worker computation delay.

\subsection{Shifted exponential distribution under server dependent scaling model}
Under this model, the computation time at a worker for evaluating $l$ gradients is given by $Y = \Delta + l.X$, where $X\sim \text{S-Exp}(\Gamma,W)$, so that $Y\sim \text{S-Exp}(\Delta+l\Gamma, lW)$. For Schemes $1$ and $2$, the number of gradients to be computed by each worker is given by $d/n$ and $2d/n$ respectively. The corresponding expected overall computation delays are given by $\Delta+\frac{\Gamma d}{n}+\frac{W.d}{n}(H_{n}-H_{s_1})  \approx \Delta+\frac{\Gamma.d}{n} + \frac{W.d}{n} \log (n/s_1)$ and $\Delta+\frac{2\Gamma d}{n}+\frac{2Wd}{n}(H_{n}-H_{s_2}) \approx \Delta+\frac{2\Gamma d}{n}+\frac{2Wd}{n} \log (n/s_2)$, where recall that $s_1 = n(1-\alpha)$ and $s_2 \approx n\sqrt{(1-\alpha)}$ (for $n$ large enough). Under this model of worker delays, the expected overall computation delay for the first scheme is smaller than the second on using $s_1\sim n {(1-\alpha)}$ and $s_2\sim n \sqrt{(1-\alpha)}$.

The above analysis highlights that there exist broad parameter regimes under various stochastic models, wherein our proposed scheme provides   improved performance over an uncoded policy in terms of the expected total computation delay. While we have only compared against a $2$-replication policy above, similar analysis can be carried out for the various other policies proposed in the previous sections. We will also supplement the above results via simulations in the next section.

%% file: simulations.tex
\section{Simulation results}
\label{Sec:simulations}
In this section, we conduct numerical simulations to compare the performance of the various schemes presented in this paper and also other approximate gradient coding schemes proposed in the literature. We consider the task of classifying images of hand-written digits from $0$ to $9$, obtained from the MNIST dataset {\cite{MNIST_data}}. We train a two layered neural network with 25 nodes in the hidden middle layer using a training set consisting of $20,000$ labelled images. We use gradient descent (with a fixed step size of {$2e-3$}) combined with cross entropy loss to iteratively optimize the weights of the neural network. 

We consider a distributed master-workers implementation for each iteration of gradient descent wherein we divide the entire training dataset into $k$ data partitions, which are then assigned to $n$ workers according to different schemes. For each scheme, every worker is designated to compute the gradient over a certain number of data partitions. We use the data-dependent scaling model described in the previous section for the worker computation time, details are discussed below. Depending on the scheme, the master waits for a certain number $n - s$ of workers to complete their computation and communicate messages to the master, after which it computes the desired gradient, updates all the weights in the neural network, and then moves to the next iteration. Thus, the time required for any iteration will be computed as the time taken for the first $n-s$ workers to finish their computation. To compare the performance of various schemes, we measure the test accuracy (over a test dataset of size $4000$) as a function of time. 

In addition to the various $(\alpha, s)$-feasible schemes discussed in the previous sections, we also compare with the following schemes from the literature:

\remove{\color{red}In schemes 2 and 3, might be worth mentioning how many workers does the master wait for, as a function of the parameter Comment SS-    Addressed as we discussed last week.}
\begin{itemize}
\item {\textit{Uncoded scheme (Forget-$s$)}}: This corresponds to a scheme where each worker is assigned one data partition and the master waits for $n-s$ workers to compute and communicate their gradient. The master computes a partial gradient by summing the gradients communicated by the workers, which corresponds to $\alpha= (n-s)/n$ in our framework. This scheme closely relates to the `fastest-k' scheme described in \cite{DBLP:journals/corr/abs-2002-11005}.
    \item \textit{$d$-Fractional Repetition Code(FRC)}: This scheme introduced in \cite{wang2019fundamental} divides $n$ workers into $d$ disjoint groups with $n / d$ workers in every group. We split the $n$ data-subsets across the workers in a group, assigning ${d}$ data-subsets to every worker in a group. All the groups are replicas of one another. Note that we choose $d=\max\Bigl\{1,\frac{
\log(n \log (\frac{n}{s}))}{\log (\frac{n}{s})}\Bigr\}$ as described in \cite[Theorem 4]{wang2019fundamental}.
    \item \textit{$\epsilon$- Batch Raptor Code(BRC)}: This scheme introduced in \cite{wang2019fundamental} assigns the data-subsets to the workers as per the Batch Raptor Code with $\epsilon$ denoting $\ell_2$-error between the recovered gradient and the actual gradient. The number of data-subsets assigned to each worker is sampled according to some degree distribution which is designed according to parameter $\epsilon$ described in \cite[Theorem 6]{wang2019fundamental}. 
    \item {\textit{Conventional Gradient Coding (CGC)}}: This scheme corresponds to $\alpha = 1$, i.e., full gradient recovery. The scheme was described and introduced in {\cite{tandon2017gradient,raviv2020gradient}} where $s+1$ data-partitions are assigned to every worker. Every worker transmits a linear combination of the corresponding gradients such that the master can compute the full gradient even if at most $s$ workers fail.
 
 \end{itemize}

For each iteration, we will use the data-dependent scaling model \cite{9174030} described in the previous section to model the computation time needed by each worker. In particular, the computation time in an iteration for a worker assigned $z$ data points is given by $z.\Delta+X$, where $\Delta$ is the fixed time per gradient computation, chosen to be $5e-7$ here\footnote{The deterministic component of the computation time for uncoded scheme and the full recovery scheme is 1e-4 and 2e-3 respectively.}; and $X$ is a random delay sampled from a Pareto distribution with parameters $(0.001,1.1)$. To model the dependency of computation times across iterations, we assume that the gradient computation time for each worker remains fixed for $300$ iterations and is independently sampled for every worker thereafter. A similar model was also used in \cite{9081964} to model dependency between stragglers across iterations. 

\subsection{Fixed $s$}
Here, we compare the performance of different schemes, each of which is designed to deal with the same number of straggler nodes $s$. In Fig~\ref{decay_n_100_s_19}, we consider $n=100$ workers and plot the test accuracy as a function of time for various schemes, each waiting for the fastest $81$ workers in each round for gradient computation; or equivalently, assuming $s=19$ stragglers in every round. We compare the performance of the cyclic schemes presented in Section~\ref{cyclic_scheme_section} (Theorem~\ref{cyclic_constr_comm_1}) for various values of $\alpha$, along with Forget-$s$ (corresponds to $\alpha = (n-s)/n = .81)$, $d$-FRC, $\epsilon$-BRC, and CGC (corresponds to $\alpha = 1$). Amongst all the schemes, except $d$-FRC, we find that our proposed cyclic scheme with $\alpha = .82$ performs the best. While $d$-FRC performs slightly better here (and in general for randomly chosen stragglers), based on the repetition structure of its computation assignment it is not difficult to construct straggler patterns where this scheme performs poorly; for example if the straggle were consecutive workers. On the other hand, our proposed scheme is designed to work with any configuration of straggler workers. 

\subsection{Fixed $\alpha$}
\begin{figure}[h]
	\centering
	\begin{minipage}{.5\textwidth}
		\centering
		\includegraphics[width=0.8\linewidth, height=0.2\textheight]{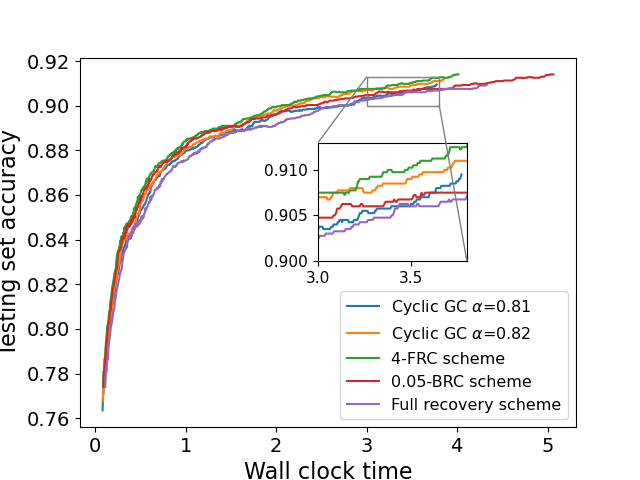}
		\caption{Test accuracy with wall -clock time for $n=100$ and $s=19$ }
		\label{decay_n_100_s_19}
	\end{minipage}%
        \centering
	\begin{minipage}{.5\textwidth}
		\centering
		\includegraphics[width=0.8\linewidth, height=0.2\textheight]{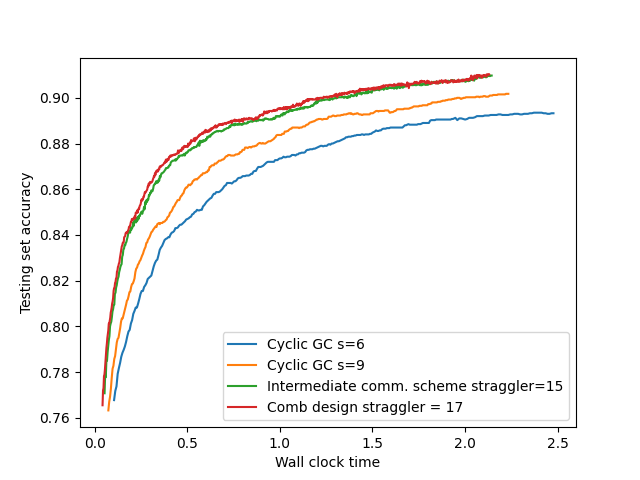}
		\caption{Test accuracy with wall-clock time for $n=30$ and $\alpha=0.8$ }
		\label{decay_n_30_alpha_8}
	\end{minipage}%
\end{figure}

All the schemes compared above had a communication load of $1$. Next, we fix the recovery fraction $\alpha$ and compare the performance of the various schemes proposed in this work. These include the combinatorial scheme described in Section~\ref{compload_ub_section} (Theorem~\ref{compload_ub}) with the least possible computation load and very high communication load, the cyclic schemes presented in Section~\ref{cyclic_scheme_section} (Theorem~\ref{cyclic_constr_comm_1}) with communication load $1$, and the schemes with intermediate communication load (described in Section~\ref{compload_inter_points}). In Fig~ \ref{decay_n_30_alpha_8}, we fix the number of workers to $n=30$ and the recovery fraction to $\alpha=0.8$. Here, the combinatorial scheme has a communication load of $435$ while the intermediate scheme has a much lower communication cost of $54$. We can see that while the combinatorial scheme demonstrates the best performance, the intermediate scheme has very similar performance while having a much lower communication load. The cyclic schemes with the lowest communication load have poorer performance, thus demonstrating a communication load-accuracy tradeoff. 


%% file: discussion.tex
\section{Discussion}
\label{Sec:Discussion}
For the exact gradient coding setup, it is known that the minimum computation load of $(s+1)/n$ and the minimum communication load of $1$ can be achieved simultaneously \cite{tandon2017gradient}. a) For the partial gradient recovery framework discussed here, while we have shown that unlike exact gradient recovery there exists a trade-off between the computation and communication loads \remove{when restricting to cyclic schemes }{which use a number of data subsets equal to $n$, in general the question remains open}. {b) While we have provided some designs for gradient coding schemes with intermediate communication and computation costs, it is not clear how close or far they are from the optimal. Also, deriving general lower bounds on the communication cost for any given partial recovery fraction remains an open problem}. c) Further, we assume each transmission message to be of a fixed size (gradient dimension $p$) and define communication load as the number of messages transmitted by each worker. Allowing splitting of gradient vectors and coding across their components as done in \cite{ye2018communication} to reduce the individual message sizes is another direction for future work. {d) The full gradient recovery problem as studied in \cite{tandon2017gradient} has a high fidelity gradient but at a high computational cost whereas the partial gradient recovery in the un-coded case has a low fidelity gradient but at a low computational cost. We believe that the gradient coding approach explored by us fits between the two extremes and we plan to conduct extensive simulations to verify the same.}

\section{Acknowledgement}

Nikhil Karamchandani acknowledges initial discussions on this problem with Bikash Kumar Dey and Chinmay Gurjarpadhye at Indian Institute of Technology, Bombay.\\


%% file: decr_claim_proof.tex
\section{Proof details of Claim \ref{decr_claim}}{\label{final_lb_details}}
We restate and prove the claim \ref{decr_claim} used in the proof of Theorem~\ref{final_lb}. 

\begin{claim*}
Consider any collection of $t$ positive integers $\{a_i\}_{1 \leq i \leq t}$. Define $a=\floor{\frac{\sum_{i=1}^{t} a_i}{t}}$ and let $t_1$ be the unique positive integer satisfying $\sum a_i = t_1.a+(t-t_1)(a+1)$. Then we have 
$\sum_{i=1}^{t}{{a_i \choose r}} \geq t_1.{{a \choose r}} + (t-t_1).{{a+1 \choose r}}$. 
\end{claim*}
 
Let us state the claim that we use to prove theorem~\ref{final_lb}.

\begin{proof}
 \begin{align*}
    & {{x +m_1 \choose r}}-{{x \choose r}}
    = \sum_{i=1}^{m_1} {{x+i-1 \choose r-1}}
 \end{align*}
 
 \begin{align*}
    & {{x+1  \choose r}}-{{x-m_2+1 \choose r}}
    = \sum_{i=1}^{m_2} {{x-i+1 \choose r-1}}
 \end{align*}
 
 These follow from ${{n \choose r}}+{{n \choose r+1}}={{n+1 \choose r+1}}$. Since ${{x+i-1 \choose r-1}}\geq {{x-j+1 \choose r-1}}$ for any $0\leq i\leq m_1$, $0\leq j\leq m_2$, we can say $\frac{{{x +m_1+1 \choose r}}-{{x+1 \choose r}}}{m_1}\geq \frac{{{x +m_1 \choose r}}-{{x \choose r}}}{m_1} \geq  \frac{{{x+1 \choose r}}-{{x-m_2+1 \choose r}}}{m_2} \geq \frac{{{x \choose r}}-{{x-m_2 \choose r}}}{m_2}$
 
 Choose the list $I$ as follows: $\{i \in [t]: a_i> a+1\}$ and list $J$ as $\{i \in [t]: a_i< a\}$. Now choose a partition of $I$ s.t $I=I_1\cup I_2$ and $I_1\cap I_2=\Phi$ and $J$ s.t $J=J_1\cup J_2$ and $J_1\cap J_2=\Phi$ s.t $|I_1\cup J_1\cup \{i \in [t]: a_i= a+1\}| = t-t_1 $. This would imply $|I_2\cup J_2\cup \{i \in [t]: a_i= a\}| = t_1 $
 
 Now denote $k_{min} = \min\limits_{m_1>1}\frac{{{x +m_1 \choose r}}-{{x \choose r}}}{m_1}$ and $k_{max}=\max\limits_{m_2>1}\frac{{{x+1 \choose r}}-{{x-m_2+1 \choose r}}}{m_2}$, thus $k_{min} \geq k_{max}$ 
 
 Thus, 
 \begin{align*}
     & \sum_{i \in I_1} \Bigl[{{a_i \choose r}} - {{a+1 \choose r}}\Bigr]+ \sum_{i \in I_2} \Bigl[{{a_i \choose r}} - {{a \choose r}}\Bigr]\\
     \overset{(a)}{\geq} & \sum_{i \in I_1} \Bigl[{{a_i-1 \choose r}} - {{a \choose r}}\Bigr]+ \sum_{i \in I_2} \Bigl[{{a_i \choose r}} - {{a \choose r}}\Bigr]\\
     \geq & k_{min} (\sum_{i \in I_1} (a_i-a-1) + \sum_{i \in I_2} (a_i-a)) 
 \end{align*}
 
 \begin{align*}
     & \sum_{i \in J_1} [{{a+1 \choose r}}-{{a_i \choose r}}]+ \sum_{i \in J_2} [{{a \choose r}} - {{a_i \choose r}}]\\
     \overset{(b)}{\leq} & \sum_{i \in J_1} [{{a \choose r}} - {{a_i-1 \choose r}}]+ \sum_{i \in J_2} [{{a \choose r}} - {{a_i \choose r}}]\\
     \leq & k_{max} (\sum_{i \in J_1} (a-a_i+1) + \sum_{i \in J_2} (a-a_i)) 
 \end{align*}
 
  Note $(a)$ follows from the fact that ${{x +m_1+1 \choose r}}-{{x+1 \choose r}} \geq {{x +m_1 \choose r}}-{{x \choose r}}$ 
and $(b)$ follows using similar reasoning. 
 
 Now 
 
 \begin{align*}
    & \sum_{i} a_i = t_1.a+(t-t_1).(a+1) \\
   \overset{(a)}{\implies} &  \sum_{i \in I_1} a_i + \sum_{i \in I_2} a_i + \sum_{i \in J_1} a_i + \sum_{i \in J_2} a_i = a.(|I_2|+|J_2|)+(a+1).(|I_1|+|J_1|)\\
   \implies & \sum_{i \in I_2} (a_i-a) + \sum_{i \in I_1} (a_i-a-1) = \sum_{i \in J_2} (a-a_i) + \sum_{i \in J_1} (a+1-a_1)
 \end{align*}
 
 Note that $(a)$ follows from the fact that $|I_1\cup J_1\cup \{i \in [t]: a_i= a+1\}| = t-t_1$ and $|I_2\cup J_2\cup \{i \in [t]: a_i= a\}| = t_1$
 Since we prove previously that $k_{min} \geq k_{max}$, we argue that 
 
 \begin{align*}
    & \sum_{i \in I_1} \Bigl[{{a_i \choose r}} - {{a+1 \choose r}}\Bigr]+ \sum_{i \in I_2} \Bigl[{{a_i \choose r}} - {{a \choose r}}\Bigr] \geq  \sum_{i \in J_1} [{{a+1 \choose r}}-{{a_i \choose r}}]+ \sum_{i \in J_2} [{{a \choose r}} - {{a_i \choose r}}]\\
     \implies & \sum_{i \in I_1\cup I_2\cup J_1\cup J_2} {{a_i \choose r}} \geq |I_1+J_1|.{{a+1 \choose r}} + |I_2+J_2|.{{a \choose r}}\\
     \overset{(a)}{\implies} & \sum_i {{a_i \choose r}} \geq t_1.{{a \choose r}} + (t-t_1).{{a +1 \choose r}}
 \end{align*}
 
 Note $(a)$ follows from the fact that $|I_1\cup J_1\cup \{i \in [t]: a_i= a+1\}| = t-t_1$ and $|I_2\cup J_2\cup \{i \in [t]: a_i= a\}| = t_1$ and definitions of $I$ and $J$.

 
\end{proof}

\remove{

Now define $b=\floor{\frac{\sum_{i=1}^{k} y_i}{k}}$ and $k_1=(b+1)k-\sum_{i=1}^{k} y_i$, thus from the claim \ref{decr_claim}, $k_1{{n-b \choose n-s}}+(k-k_1){{n-b-1 \choose n-s}} \leq \sum_{i=1}^{k} {{n-y_i \choose n-s}}$ since $n-b-1=\floor{\frac{\sum_{i=1}^{k} (n-y_i)}{k}}$ and $\sum_{i=1}^{k} (n-y_i) = (k-k_1).(n-b-1)+k_1.(n-b)$. The LHS is the smallest when $\sum_{i=1}^{k}{y_i}=n.k.l$ since $b$ increases with increase in $\sum{y_i}$ and $k_1$ decreases with $\sum{y_i}$ when $b=\floor{\frac{\sum_{i=1}^{k} y_i}{k}}$ remains unchanged. Thus, the inequality reduces to $k{{n-b-1 \choose n-s}}\leq  \sum_{i=1}^{k} {{n-y_i \choose n-s}}\leq {{n \choose s}}k (1-\alpha)$ where $b=\floor{{n.l}}$ because ${{n-a-1 \choose n-s}} \leq {{n-a \choose n-s}}$ which proves Theorem \ref{final_lb}.

We restate Claim \ref{decr_claim} and prove it below.

\begin{claim*}
The following claim holds true for any set of positive integers $\{a_i\}_{1 \leq i \leq t}$. Define $a=\floor{\frac{\sum_{i=1}^{t} a_i}{t}}$ and $t_1$ is a unique positive integer satisfying $\sum a_i = t_1.a+(t-t_1)(a+1)$. Then we have 
$\sum_{i=1}^{t}{{a_i \choose r}} \geq t_1.{{a \choose r}} + (t-t_1).{{a+1 \choose r}}$. 
\end{claim*}
}

%% file: constr_lower_comm.tex
\section{Proof details of Theorem \ref{impr_comm_cost}}{\label{proof_details_improved_comm}}

\subsection{Construction}

Let us re-index each data subset by the set of worker indices it is assigned to. For example, data subset $D_{T_J}$ is assigned to workers indexed by set $J$. Consider all possible subsets containing 1 and order them in lexicographic order for example sets $\{\{1,2,3\},\{1,2,4\},\{1,2,5\},\{1,2,6\},\{1,3,4\},\{1,3,5\}\}$ are some subsets of cardinality 3 sorted in lexicographic order. Two sets $I$ and $J$ s.t $I,J \in [n]$ differ by a cyclic shift if $J=\{1+(y+a-1)\%n| y \in I\}$ for some $a \in [n]$.

Consider the first distinct $\frac{1}{y}{{n-1 \choose y-1}}$ subsets of cardinality $y$ of $[n]$ with each subset containing 1 such that no two subsets differ by any cyclic shift. Let us denote this collection of sets by $P_1$. Now choose worker $W_1$ corresponding to all data subsets $D_{T_J}$ $\forall J \in P_1$. Thus, in other words $t_{T_J}=1$ $\forall J \in P_1$. Recall from the proof of Theorem~\ref{impr_comm_cost} that worker $W_{t_j}$ is chosen for data subset $D_j$ such that $t_j$ belongs to the indices of the $j$ workers to whom data subset $D_j$ is assigned.  

Now let us define the set $P_2$. Increment each element in every subset of $P_1$ by 1 with rollover to 1 if crosses $n$ to obtain $P_2$. Formally we denote $P_2=\{\{1+(u \mod n)| u \in J\}| J \in P_1\}$. Similarly we choose worker $W_2$ corresponding to all data subsets $D_{T_J}$ $\forall J \in P_2$. In general, we define the subset $P_x$ $\forall x \in [n]$ by increasing each element of $P_1$ by $x-1$ with rollover to 1 if the sum crosses $n$. Formally we denote $P_x=\{\{x-1+(u \mod n)| u \in J\}| J \in P_1\}$. Similarly we choose worker $W_x$ corresponding to all data subsets $D_{T_J}$ $\forall J \in P_x$.

Let us work out an example for the case of $n=7$ and $y=4$. Note that $n$ and $y$ are co-prime. 

\begin{example}

The distinct ${{n-1 \choose y-1}} = 20$ subsets in lexicographic order containing 1 can be written as $\{\{1,2,3,4\},\{1,2,3,5\},\\ \{1,2,3,6\},\{1,2,3,7\},\{1,3,4,5\},...\{1,5,6,7\}\}$. However observe that the subsets $\{1,2,3,4\}$ and $\{1,2,3,7\}$ differ by a cyclic shift of 6 and the subsets $\{1,2,3,6\}$ and $\{1,3,4,5\}$ also differ by a cyclic shift of $5$. Thus we choose the set $P_1$ as $\{\{1,2,3,4\},\{1,2,3,5\},\{1,2,3,6\},\{1,3,4,6\},\{1,3,4,7\}\}$, similarly we define the set $P_2$ as $\{\{2,3,4,5\},\{2,3,4,6\},\{2,3,4,7\}\\ ,\{2,4,5,7\} ,\{2,4,5,1\}\}$ by incrementing each element in each set of $P_1$ by 1 and thus set $P_7$ is defined as $\{\{7,1,2,3\},\{7,1,2,4\}\\ ,\{7,1,2,5\},\{7,2,3,5\},\{7,2,3,6\}\}$ and hence we choose worker $W_1$ corresponding to data subsets which are assigned to the set of workers indexed by sets in $P_1$. Similarly we choose worker $W_2$ corresponding to data subsets which are assigned to the set of workers indexed by sets in $P_2$ and so till sets in $P_7$. 
\end{example}

\subsection{Proof that the above construction works}

Clearly no two subsets in any set $P_i$ can be identical since they have been obtained by incrementing every element in distinct subsets in $P_1$ by $i-1$. 

We first show that there can be no element in both $P_i$ and $P_j$ for $i \neq j$. Let us prove it by contradiction by assuming that there exists a subset $J$ in both $P_i$ and $P_j$. Suppose $J$ was obtained in set $P_i$ by shifting elements of subset $A$ in $P_1$ by $i-1$ and $J$ was obtained in set $P_j$ by shifting elements of another subset $B$ by $j-1$. Thus, subset $B$ can be obtained from $A$ by shifting each element of $A$ by $j-i$ which is a contradiction since both $A$ and $B$ are consecutive distinct lexicographic subsets containing 1. 

Let us consider the other case if subset $J$ is obtained in set $P_i$ and set $P_j$ by shifting elements of the same subset $A$ by $i-1$ and $j-1$ respectively. Thus shifting elements of subset $J$ by $j-i$ gives the same subset. Consider the smallest element $a$ such that shifting elements of $J$ by $a-1$ gives the same subset $J$. Suppose the $(1+(t+r-1)\% y)^{th}$ element of $J$ be equal to $t^{th}$ element of the list obtained after shifting elements of $J$ by $a$ $\forall t \in [n]$. Let us denote the elements of $J$ as $[j_1,j_2,..j_y]$ where $j_1 < j_2 < ..< j_y$. This would imply that $\sum_{z=1}^{r}j_{1+(t+z-1)\%r}=a-1$ for every integer $t$ i.e. any consecutive set of $r$ elements element has the sum to be $a-1$. 

Suppose $y$ is not a multiple of $r$. Suppose not and say the remainder when $r$ divides $y$ is given by $q$, then we can argue that $\sum_{z=1}^{q}j_{1+(z+t-1)\%r}$ remains the same for all $t$ clearly the sum of which is smaller than $a-1$, thus there would exist an integer smaller than $a$ say $b$ such that shifting elements of $J$ by $b$ gives the same subset $J$.

Thus $y$ is a multiple of $r$ say $y=r.m$, hence $n = \sum_{z=1}^{y}j_z=m.\sum_{z=1}^{r} j_z=m.(a-1)$ since any the sum of any set of $r$ consecutive elements remain the same. Thus $n$ and $y$ have the same factor implying they are not co-prime. 

Thus, we proved that no two elements of two distinct sets $P_i$ and $P_j$ can be the same. Since the sum of cardinalities of all subsets $\{P_i\}$ is $\frac{n}{y}{{n-1 \choose y-1 }}={{n \choose y}}$ implying that each worker is chosen exactly the same no of times i.e $\frac{1}{y}{{n-1 \choose y-1}}$ and there is a worker chosen for every data subset.

%% file: proof_cyclic_constr_1.tex
\section{Correctness argument of the construction proposed in Theorem \ref{cyclic_constr_comm_1}}\label{cyclic_constr_comm_1_proof}

Recall the construction from the proof of Theorem \ref{cyclic_constr_comm_1} where $W_i$ contains the data subsets $D_i,D_{i+1},..D_{1+(i+r-2)\%r}$ where $r=s+1+\beta-n$. \remove{This follows from the definition of cyclic schemes. }Let the workers denoted by $W_1,...,W_{\beta}$ be grouped into $r$ groups
each group containing $\frac{\beta}{r}$ workers. Let us denote the groups by $\{\mathcal{A}_j\}_{j \in [r]}$. Suppose group $\mathcal{A}_j$ contains the $\frac{\beta}{r}$ workers $W_j,W_{j+r},...W_{j+\beta-r}$ which ensures that no one worker is present in two different groups. Note that worker denoted by $W_i$ belongs in group $\mathcal{A}_{f(i)}$ where $f(i)=1+(i-1)\%r$.

Consider the set of straggling workers be denoted by $S$ s.t $|S|=s$. Suppose there exists a group $\mathcal{A}_i$ with no straggling worker present, this would imply the existence of $\frac{\beta}{r}$ workers with disjoint set of data subsets implying the master would be able to calculate the sum of $\beta$ gradients from the results computed by the non-straggling workers. Suppose there does not exist a group $\mathcal{A}_i$ without any straggling worker present in it i.e each group has at least one straggling worker present.

\begin{algorithm}{\label{back_move_straggler}}
\SetAlgoLined

Choose largest $i \in [\beta]$ s.t $i \in S$.\\

\While{$\exists$ $v<i$ s.t $W_v \in \mathcal{A}_{f(i)}$\text { and }$W_v \in [S]$}{

 Choose largest $j<i$ s.t  $W_j \in S$.\\
 \remove{
 \eIf{$\exists k$ s.t $j<k<i$ and $W_k \in [S]$}{
 
 Choose largest $k$ s.t $j<k<i$ and $W_k \in [S]$.\\
 i =k.\\
 }
 {
   i =j\\
  }}
  i =j \\
 
}

Output $i$.

\caption{Stopping Straggler 1}
\end{algorithm}

Note that worker $W_i$ returned from Alg~\ref{back_move_straggler} would have no worker $W_j$ in $\mathcal{A}_{f(i)}$ s.t $j<i$ and $W_j \in S$. Now consider the entire set of groups visited in the algorithm described as $I$. We can argue if $I=x+1$, there must exist at least $x$ workers $W_j$ satisfying $j<i$ and $W_j \in S$ (at least one worker from each of the $x$ rows). Also note that there must be at least one straggling worker corresponding to each of $(r-x-1)$ groups which were not visited in the algorithm since each group has at least one straggling worker. Note that each of these workers must have its index smaller than $i$, thus there would exist at least $(r-x-1)+x=r-1$ workers behind $W_i$. Now consider worker $W_i$ in group $\mathcal{A}_{f(i)}$ and suppose we have $m$ workers $W_j$ s.t. $j\geq i$ and $W_j \in \mathcal{A}_{f(i)}$. Suppose we denote all the workers in group $\mathcal{A}_{f(i)}$ as $W_{i_1},...,W_{i_{\frac{\beta}{r}}}$, thus $i=i_{\frac{\beta}{r}-m+1}=f(i)+(\frac{\beta}{r}-m)r$. 
We now claim there must exist at least a vector of integers $[k_1,k_2,...,k_m]$ with these indices lying in the set of indices of non-straggling workers satisfying
\begin{align}{\label{condn1}}
    & k_t-k_{t-1}\geq r \forall t \in [m-1], \text{ }, k_1\geq i\nonumber\\
    & k_m \leq n+f(i)-r\remove{=n-\beta+(f(i)+\beta-r)=n-\beta+(i)+(f(i)+\beta-r-i)=i+(n-\beta)+(m-1).i}
\end{align}
Note that the conditions mentioned above would ensure that no overlap between the data subsets assigned to workers $W_{i_1},W_{i_2},...,W_{i_{\frac{\beta}{r}-m}},W_{k_1},...,W_{k_m}$. 
The minimum size of set $Z$ s.t there is no solution of $\{k_v\}_{v \in [m]}$ satisfying $k_v \notin Z \forall v \in [m]$ and \eqref{condn1} is given by $(n+f(i)-r)-i-(m-1).r+1=n+f(i)-m\times r-i+1= n + f(i)-m\times r -f(i)-(\frac{\beta}{r}-m)r +1= n-\beta+1$. However, the number of straggled workers $W_j$ s.t $j \geq i$ is at most $s-r+1=n-\beta$ which would imply that there exists at least a vector $[k_1,...,k_m]$ satisfying \eqref{condn1}, showing the existence of a set of $\frac{\beta}{r}$ non-straggling workers such that the data subsets assigned to them don't overlap.

\remove{
$y_m \leq (m-1)r+(n-\beta)$ given that $y_t-y_{t-1} \geq r$ and $y_i \notin Z \forall i \in [m]$ since $n+f(i)-r=n-\beta+(f(i)+\beta-r)=n-\beta+(f(i)+\beta-m\times r)+(m-1)r=i_t+(n-\beta)+(m-1)r$. The minimum size of set $Z$ which can be shown to be $n-\beta+1$.
}
\remove{
Consider the number of straggled workers behind $W_i$, clearly there are at least $(r-x-1)$ groups which are not visited in the algorithm above, each of which must have a straggled worker behind $W_i$. Thus the total number of straggled worker behind $W_i$ is at least $(x)+(r-x-1)=r-1$ (the first term corresponds to workers in visited groups, the second not visited groups)
}

We give two examples of sets of stragglers to demonstrate the proof strategy mentioned above and describe two sets of stragglers for $n=18$ and $\beta=15$ and $s=7$
\begin{example}
Recall that the set of workers is given by $W_1,W_2,...,W_{18}$. Suppose the set of straggling workers is denoted $W_{15}$, $W_{13}$, $W_{10}$, $W_8$, $W_4$, $W_{12}$ and $W_{11}$. Note that the group $\mathcal{A}_1$ would contain the workers $W_1$, $W_6$ and $W_{11}$, the group $\mathcal{A}_2$ would contain the workers $W_2$, $W_6$ and $W_{12}$ and so on and finally the group $\mathcal{A}_5$ contains the workers $W_5$, $W_{10}$ and $W_{15}$. First it is important to note that each group has at least one straggling worker.
 
Note that the largest indexed straggling worker amongst the first fifteen is 15. Thus $i$ is initialised to 15. However there is a straggling worker with a smaller index $10$ in the same group $\mathcal{A}_5$, hence we set $i$ get to 13. Again, since there is another straggling worker with index $8$ in the group $\mathcal{A}_3$, we set $i$ as 12. However, since there is no straggling worker with smaller index in group $\mathcal{A}_2$, we return 12 Note that $W_{12}$ has exactly 4 stragglers with indices smaller than 12 none of them being in the group. Note that $m$ is 1 in this case and we can choose $k_1$ as 14 (belonging to set of non-straggling workers) satisfying \eqref{condn1}. Thus we obtain the can use the transmissions by workers $W_2$,$W_7$ and $W_{14}$ to obtain the sum of 15 gradients.
\end{example}

\begin{example}
Recall that the set of workers is given by $W_1,W_2,...,W_{18}$. Suppose the set of straggling workers is denoted $W_{14}$, $W_{13}$, $W_{9}$, $W_8$, $W_6$, $W_7$ and $W_5$. Note that the group $\mathcal{A}_1$ would contain the workers $W_1$, $W_6$ and $W_{11}$, the group $\mathcal{A}_2$ would contain the workers $W_2$, $W_6$ and $W_{12}$ and so on and finally the group $\mathcal{A}_5$ contains the workers $W_5$, $W_{10}$ and $W_{15}$. Also observe that each group has at least one straggling worker.
 
Note that the largest indexed straggling worker amongst the first fifteen is 14. Thus $i$ is initialised to 14. However there is a straggling worker with a smaller index 9 in the same group $\mathcal{A}_4$, hence we set $i$ get to 13. Again, since there is another straggling worker with index $8$ in the group $\mathcal{A}_3$, we set $i$ as 8 as there is no straggling worker with index larger than 8. Note that $W_{8}$ has exactly 4 stragglers with indices smaller than 8 with none of them being in the group $\mathcal{A}_3$. Note that $m$ equals 2 in this case and we can choose $k_1=10$ and $k_2=15$ (belonging to set of non-straggling workers) satisfying \eqref{condn1}. Thus we obtain the can use the transmissions by workers $W_3$,$W_{10}$ and $W_{15}$ to obtain the sum of 15 gradients. 

\remove{
. However, since there is no straggling worker with smaller index in group $\mathcal{A}_2$, we return 12 Note that $W_{12}$ has exactly 4 stragglers with indices smaller than 12 none of them being in the group. Note that $m$ is 1 in this case and we can choose $k_1$ as 14 satisfying \ref{condn1}. Thus we obtain the can use the transmissions by workers $W_2$,$W_7$ and $W_{14}$ to obtain the sum of 15 gradients.

}
\end{example}


%% file: proof_cyclic_constr_2.tex
\section{Correctness argument of the construction proposed in Theorem \ref{cyclic_constr_comm_2}}\label{cyclic_constr_comm_2_proof}

Recall that worker $W_i$ contains the data subsets $D_i,D_{i+1},..D_{1+(i+r-2)\%r}$ where $r=s+1+\beta-n$. \remove{This follows from the definition of cyclic schemes. } Note that we denote the remainder by $x$ when $r$ divides $\beta$ i.e $x = (\beta \mod r)$. Let the workers denoted by $W_1,...,W_{\gamma}$ be grouped into $r$ groups
each group containing $\frac{\gamma}{r}$ workers where $\gamma=\beta-x$ which is clearly divisible by $r$. Let us denote the groups by $\{\mathcal{A}_j\}_{j \in [r]}$. Suppose group $\mathcal{A}_j$ contains the $\frac{\gamma}{r}$ workers $W_j,W_{j+r},...W_{j+\gamma-r}$ which ensures that no one worker is present in two different groups. Note that worker denoted by $W_i$ belongs in group $\mathcal{A}_{f(i)}$ where $f(i)=1+(i-1)\%r$.

Consider the set of straggling workers be denoted by $S$ s.t $|S|=s$. 

\begin{algorithm}{\label{back_move_straggler_2}}
\SetAlgoLined

Choose largest $i \in [\gamma]$ s.t $i \in S$.\\

\While{$\exists$ $k<i$ s.t $W_k \in \mathcal{A}_{f(i)}$ and $W_k \in [S]$}{

 Choose largest $j<i$ s.t. $W_j \in S$.\\
 \remove{
 \eIf{$\exists k$ s.t $j<k<i$ and $W_k \in [S]$}{
 
 Choose largest $k$ s.t $j<k<i$ and $W_k \in [S]$.\\
 i =k.\\
 }
 {
   i =j\\
  }}
  i=j \\
 
}

Output $i$.

\caption{Stopping Straggler 2}
\end{algorithm}

Note that worker $W_i$ returned from the Algorithm \ref{back_move_straggler_2} would have no worker $W_j$ in $\mathcal{A}_{f(i)}$ s.t $j<i$ and $W_j \in S$.

Now consider the entire set of groups visited in the algorithm described as $I$. We can argue if $I=z+1$, there must exist at least $z$ workers $W_j$ satisfying $j<i$ and $W_j \in S$ (at least one worker from each of the $x$ rows). Suppose there exists $m$ workers $W_j$ s.t $j \geq i$ and  $W_j \in \mathcal{A}_{f(i)}$ and we denote all the workers in group $\mathcal{A}_{f(i)}$ as $W_{i_1},...,W_{i_{\frac{\gamma}{r}}}$, thus $i=i_{\frac{\gamma}{r}-m+1}=f(i)+(\frac{\gamma}{r}-m)r$. We consider two cases i.e. when each group $\{\mathcal{A}_i\}$ has at least one-straggling worker and when there exist groups without any straggling worker.

\textbf{{\label{first_case_thm_6}}
Case I: There does not exist any group without any straggling worker.
}\\
Since the algorithm visited exactly $(z+1)$ distinct groups implying the existence of exactly $(r-z-1)$ groups which are not visited each of which must have at least one straggling worker with index smaller than $i$. Thus there exist at least $(r-z-1)+z=r-1$ workers with index smaller than $i$. Now consider the smallest index $u$ larger than $i$ such that $W_u$ is a non-straggling worker. 

\remove{
Now consider worker $W_i$ in group $\mathcal{A}_{f(i)}$ and suppose we have $m$ workers $W_j$ s.t. $j\geq i$ and $W_j \in \mathcal{A}_{f(i)}$. Suppose we denote all the workers in group $\mathcal{A}_{f(i)}$ as $W_{i_1},...,W_{i_{\frac{\beta}{r}}}$ where $i=i_{\frac{\gamma}{r}-m+1}=f(i)+(\frac{\gamma}{r}-m)r$
}

We now claim there must exist at least a vector $[k_1,k_2,...,k_m]$ satisfying

\begin{align}{\label{condn2}}
    & k_t-k_{t-1}\geq r, \forall t \in [m-1], k_1\geq u+x\nonumber \\
    & k_m \leq n+f(i)-r\remove{=n-\beta+(f(i)+\beta-r)=n-\beta+(i)+(f(i)+\beta-r-i)=i+(n-\beta)+(m-1).i}
\end{align}

Note that the conditions mentioned above would ensure that no overlap between the sum of all the $r$ data subsets transmitted by the workers $W_{i_1},W_{i_2},...,W_{i_{\frac{\gamma}{r}-m}},W_{k_1},...,W_{k_m}$ and the data subset of first $x$ gradients transmitted by the worker $W_{u}$. Note that a key difference in this approach is that the data subsets assigned to workers $W_{i_1},W_{i_2},...,W_{i_{\frac{\gamma}{r}-m}},W_{k_1},...,W_{k_m}$ and $W_u$ may overlapping unlike the proof of Theorem~\ref{cyclic_constr_comm_1}.

Recall from the definition of $\gamma$ that $\gamma=\beta-x$.

The minimum size of set $Z$ s.t there is no solution of $\{k_v\}_{v \in [m]}$ satisfying $k_v \notin Z \forall v \in [m]$ and \eqref{condn2} can be shown to be $n+f(i)-r-(u+x)-(m-1)r+1=n+f(i)-(u+x)-m\times r+1=(n-u)+(i-\beta)+(f(i)-i+\gamma)-m \times r+1=(n-u)+(i-\beta)+1$.

\remove{
Since $i= i_{1+\frac{\gamma}{r}-m}=f(i)+(\frac{\gamma}{r}-m)r$, it we may equivalently calculate the minimum size of set $Z$ s.t there is no solution of $\{k_v\}_{v \in [m]}$ to $y_m \leq (m-1)r+(n-\gamma)+(i_{1+\frac{\gamma}{r}-m}-(u+x))$ given that $y_t-y_{t-1} \geq r$ and $y_i \notin Z \forall i \in [m]$ since $n+f(i)-r=n-\gamma+(f(i)+\gamma-r)=n-\gamma+(f(i)+\gamma-m\times r)+(m-1)r=i+(n-\gamma)+(m-1)r$. The minimum size of set $Z$ which can be shown to be $n-\gamma+(i-(u+x))+1=n-\beta-u+i+1$.
}
However note that since $W_u$ is the smallest index non-straggling worker larger than $i$, there are at least $u-i$ straggling workers from  $W_i$ to $W_{u}$. However, since the total number of stragglers starting from $W_{u+x}$ is $(s-(r-1)-(u-i)=-\beta+n-u+i$ which is clearly smaller than the minimum size needed to ensure no solution of \eqref{condn2}.

Thus we can recover the desired sum of $\beta$ gradients from the transmission of sum of $r$ gradients by the workers $W_{i_1},W_{i_2},...,W_{i_{\frac{\gamma}{r}-m}},W_{k_1},...,W_{k_m}$ and the sum of $x$ gradients by worker $W_{u}$

\textbf{
Case -II: There exist groups $\{\mathcal{A}_j\}_{j \in [r]}$ with no straggling workers in any of these groups.
}
\remove{

\textbf{ Case II-(a): The largest and the smallest indices of groups without any straggling worker differ by at least $x$.\\} Under this assumption, we can argue that there would exist a set of $\frac{\gamma}{r}+1$ workers such that the first $x$ data-subsets of the one worker would be non-overlapping with all the data subsets of the other $\frac{\gamma}{r}$ workers.

Suppose the largest index of such groups is denoted by $L_1$ and the smallest index of such group is denoted by $L_2$. Suppose the workers of group $\mathcal{A}_{L_1}$ is denoted by $\{W_{T_1},W_{T_2},...,W_{T_{\frac{\gamma}{r}}}\}$. Note that $T_1=L_1$ and $T_i-T_{i-1}= r$. We can show that the data subsets of workers $W_{T_1},W_{T_2},...,W_{T_{\frac{\gamma}{r}}}$ and the first $x$ data subsets of $W_{L_2}$ are non-overlapping since $L_1-L_2 \geq x$, thus the master can recover the sum of $\beta$ gradients from non-straggling workers.
}

Suppose the set of all such indices of groups without any straggling workers is denoted by $J$ with $|J|=t$. Since it is a cyclic scheme and cyclic shifts in data subsets assigned to workers don't make any difference, we assume the smallest and largest index of $J$ to be $1$ and $a$ where $a\leq r$. 

{Now consider $B_w=\mathcal{A}_w \cup \{W_{w+(\gamma)}\} \forall w \in J$ and $w \geq r-x$ Note that since the largest and the smallest index of $J$ differ by atmost $r$, there can be any common worker between two distinct subsets $B_i$ and $B_j$ $i,j \in J$. Also note that $w+\gamma \leq n$ since $w \leq r$ and $r-x \leq n - \beta$ which implies $r+\gamma \leq n$ as $\gamma = \beta -x$. 
}
Suppose there exists $w \in J$ s.t no worker in $B_w$ straggles, then choose the master can recover the sum of $\beta$ gradients by the sum of $r$ gradients of workers $\mathcal{A}_w$ and the sum of first $x$ gradients transmitted by $W_{w+\gamma}$.

Suppose there does not exist any worker $w \in J$ s.t no worker in $B_w$ straggles. In this case, each worker in $W_{w+(\gamma)} \forall w \in J$ must straggle as the other workers in $\mathcal{A}_w \forall w \in J$ don't straggle. Choose the largest element in $J$ which has been assumed to be $a$ in this case and thus consider the worker $W_{a+\gamma}$. The worker with index $i$ has clearly smaller index than all workers in $W_{w+(\gamma)} \forall w \in J$. Also we know there are at least $(r-z-1)+(z-t)=(r-t-1)$ straggled workers with indices smaller than $i$. Thus we have $(r-t-1)+1+(t-1)=r-1$ straggled workers with indices smaller than $a+\gamma$. We denote the workers in group $\mathcal{A}_a$ as $\{W_{a_1},W_{a_2},...,W_{a_{\frac{\gamma}{r}}}\}$ Using a very similar argument as in Case \ref{first_case_thm_6}, we can show that there would exist $z$ satisfying $z-a_{\frac{\gamma}{r}} \geq r$ and $z \leq n+a-x$ with $W_z$ being a non-straggling worker. This can be argued from the fact the number of straggling workers with indices larger than or equal to $a_{\frac{\gamma}{r}}$ is at most $(s-r+1)=(n-\beta)$ since the number of stragglers with index less than $a_{\frac{\gamma}{r}}$ is at most $r-1$. Thus we can recover the desired sum of $\beta$ gradients from the transmission of sum of $r$ gradients by the workers $W_{a_1},W_{a_2},...,W_{a_{\frac{\gamma}{r}}}$ and the sum of $x$ gradients by worker $W_z$.

%% file: proof_lower_bound_low_comp.tex
\section{Proof of Theorem \ref{lower_bound1_low_comp}}{\label{lower_bound1_low_comp_section}}
Let us prove by contradiction. Suppose we have $l=\frac{2}{n}$ i.e. every worker transmits and computes a linear combination of gradients of at most two data subsets and represent every data subset as a node in the graph. Since every worker can be assigned at most 2 data subsets, we denote it as an edge between the two corresponding nodes if it indeed transmits a linear combination of two data subsets; else we represent it as a self-loop around the node corresponding to the data subset assigned to it. We divide the problem into two cases (described below) and prove it for each case. \remove{Also note that the second case is the most general case.}

\begin{itemize}
    \item We assume a uniform distribution of two data subsets to every worker with each data subset being assigned to at exactly two workers. Thus, the graph consists of disconnected components with each component being a cycle.
    
    \item We assume the graph consists of one or more disconnected components.
\end{itemize}

\textbf{\label{first_case}
Case-I: The graph consists of disconnected component with each component being a cycle.}

\remove{
	 Then each worker transmits a linear combination of the two gradients computed by it. Let us denote all the $n$ gradients by nodes of a graph and two nodes are connected by an edge if there exists a worker which computes both the gradients corresponding to the nodes.
}	
	
\begin{figure}[!h]
    	\centering
    	\includegraphics[scale= 0.4]{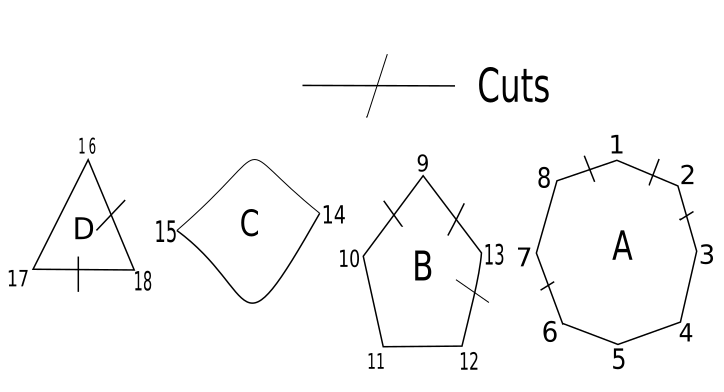}
    	\caption{Representation as  data subsets as nodes with numbers denoting nodes and letters denoting components}
    	\label{graph_representation}
    \end{figure}

	Since every gradient is being computed by two workers, each node must be present in two edges. Also note that we have $n$ edges and $n$ nodes, thus the graph must be comprised of disjoint cyclic components with each component being a cycle. There could be a pair of isolated nodes with a pair of edges connecting them which we treat as a connected cyclic component only like the component C in the Fig. \ref{graph_representation}. We denote the sizes(vertices/edges) of the components by $c_1,c_2,...c_t$ if there are $t$ such components with $\sum c_t =n$. W.L.O.G, we assume $c_1 \leq c_2 \leq c_3 ...\leq c_t$. \remove{Let $p$ denote the smallest index such that $\sum_{i=1}^{p} c_i\geq s+1$.}Note that the lines on edges denoting the straggling workers in the diagram and every edge denotes a unique worker and now state the following claim and use it for the proof of the theorem.
    
	\begin{claim} \label{claim:sum_grad}
    If the above gradient code is $(\alpha,s=n-\beta+1)$ feasible where $\beta=\ceil{n\alpha}$, then each worker must transmit the sum of the gradients of the data subsets assigned to it,   
    \end{claim}
	
	\begin{proof}
	    We divide the proof into two parts - for edges in cycles with size larger than or equal to $s$ and cycles with size smaller than $s$.\\
	    \textbf{
	    Case (a): Consider any edge (worker) in some component with size at least $s+1$.} 
	    
	    \remove{
	    
	    For simplicity, we denote the size of this component by $x$ and the node of this component is $A_1,A_2,...,A_x$ and the edges are $A_1-A_2,A_2-A_3,...A_x-A_1$. We assume the edge corresponding to this worker as $A_1-A_2$. Now we consider the case when workers corresponding to edges $A_x-A_1$,$A_2-A_3$, $A_3-A_4$,...,$A_s-A_{s+1}$ straggle- for example the cut in component $A$. In this case clearly the gradients to  data subsets $A_3,..,A_s$ won't be accessible to the master at all, thus these $s-2$ gradients cannot be present in the sum computed by the master, thus the master has access to at most $(\beta+1)$ gradients with data subsets corresponding to $A_1$ and $A_2$ being present in exactly one non straggling worker. Since the master has to compute the sum of at least $\beta$ gradients, the coefficient of  the gradients corresponding to data subsets $A_1$ and $A_2$ in the worker which transmits their linear combination has to be the same.
	    }
    
        We denote the size of this component by $x$ with the nodes of this component being $A_1,A_2,...,A_x$ in a cyclic order assuming the edge corresponding to this worker is $A_1-A_2$. Now we consider the case when workers corresponding to edges $A_x-A_1$,$A_2-A_3$, $A_3-A_4$,...,$A_s-A_{s+1}$ straggle- for example the cut in component $A$. In this case the master has access to at most $(\beta+1)$ gradients with data subsets corresponding to $A_1$ and $A_2$ being present in exactly one non straggling worker. As the master has to compute the sum of at least $\beta$ gradients, the coefficient of  the gradients corresponding to data subsets $A_1$ and $A_2$ in the worker which transmits their linear combination has to be the same.
        
    \textbf{
	    Case (b) : Consider an edge in some component $r$ of size at most $s$.
    }\\
      For simplicity, we denote the size of this component by $x$ and the node of this component is $A_1,A_2,...,A_x$ in a cyclic order. \remove{and the edges are $A_1-A_2,A_2-A_3,...A_x-A_1$.}  Let $p_{min}$ be the minimum $i$ satisfying $\sum_{j=1;j \neq r}^{i} c_i \geq s-x+1$ and $p_{last}= s-x+1-\sum_{j=1; j \neq r}^{p_{min}-1} c_i$. Now, we straggle the following $s$ workers-
    
    \begin{itemize}
    	\item All workers in the first $p_{min}-1$ components excluding component numbered $r$.
    	\item Exactly $p_{last}$ continuous edges of component $c_i$ like the cut shown in component $B$.
    	\item All edges except $A_1-A_2$ in component numbered $r$ as the cut shown in component $D$
    \end{itemize} 
     
     \remove{ 
     Now we can observe that the master won't have access to $s-x+x-2=s-2$ gradients if the above set of $s$ workers straggle. Thus the master has access to at most $(\beta+1)$ gradients with data subsets corresponding to $A_1$ and $A_2$ being present to only one non-straggling worker. Since the master has to compute the sum of at least $\beta$ gradients, the coefficient of  the gradients corresponding to data subsets $A_1$ and $A_2$ in the worker which transmits their linear combination has to be the same.
     }
     
     Now we observe that the master has access to at most $(\beta+1)$ gradients with data subsets corresponding to $A_1$ and $A_2$ being present in only one non-straggling worker {$\tilde{W}$}. As the master has to compute the sum of at least $\beta$ gradients, the coefficient of the gradients corresponding to data subsets $A_1$ and $A_2$ in non-straggling worker $\tilde{W}$ has to be the same. 
     Thus, we proved that each worker transmits the sum of the gradients of data subset assigned to it without assuming $\beta$ is odd. \end{proof}

\remove{

However, the master can not have at most $n-\beta=s-1$ missing gradients which would imply at least gradient $A_1$ or gradient $A_2$ must be computed by the master. Also note that  data subsets $A_1$ and $A_2$ is present in just one worker and the master must have coefficient of all the gradient to be the same, which would further imply that the coefficient of  data subsets $A_1$ and $A_2$ in the worker corresponding to edge $A_1-A_2$ must be the same.

} 

\remove{
Thus, we prove that the coefficients of the gradient of each data subset in each worker must be the same without assuming $\beta$ is odd.}

    \remove{
    
    However the master can have at most $n-\beta=s-1$ missing gradients in the sum computed by it. Also the gradients corresponding to the  data subsets $A_1-A_2$ is present in only worker would imply that both coefficients of the gradients in the worker must be same.
    }
    

\remove{However, if $\beta$ is odd, this construction cannot yield the desired sum of the gradients at the master.}

If $\beta$ is odd, we now choose a set of stragglers such that the master cannot compute the sum of any set of $\beta$ or more gradients.
We know that the sizes of the components are $c_1,c_2,...,c_t$ with $\sum_t c_t=n$ and choose $d_1,d_2,...,d_n$  such that $\sum_t d_t=s$ with $d_i = c_i \forall 1 \leq i < m$ for some $m$, $d_m \leq c_m$ and $d_i=0 \text{ } \forall i \geq m+1$. \remove{The selection of stragglers show that the master won't have access to at least $s-1$ gradients. None of the gradients corresponding to any of data subsets present in any of the first $m-1$ workers can be computed by the master. \remove{Also $c_m-d_m$ must clearly be even as $\sum_t (c_t-d_t)$ is even.}} Suppose $c_m-d_m$ is odd, then there there must exist some $i \geq m+1$ s.t $(c_i-d_i)$ is odd as $\sum\limits_{i \geq m+1} (c_i-d_i)$ which is odd. Thus, we argue $c_j$ to be odd for some $j \geq m+1$ as $d_j=0 \forall j \geq m+1$. Now we decrease $d_m$ by 1 and set $d_j$ to 1 ensuring $\sum d_t$ remains $s$ and define the $s$ workers which straggle. All the workers corresponding to the edges in the first $m-1$ components straggle and a set of continuous $d_m$ edges in component numbered $m$ straggle and continuous $d_j$ edges in component numbered $j$ straggle.

\begin{figure}[h]
	\centering
   \begin{minipage}{0.46\textwidth}
		\centering
			\label{Fig:a}
	\includegraphics[scale = 0.4]{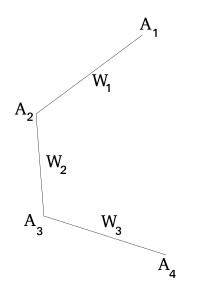} 
	\caption{A continuous set of odd number of edges}\label{fig:odd_edges}  
	\end{minipage}
	\begin{minipage}{0.46\textwidth}
   	\centering
   	\label{Fig:b}
   	\includegraphics[scale = 0.4]{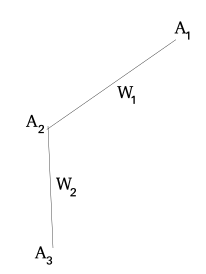}
   	\caption{A continuous set of even number of edges}\label{fig:even_edges}  
   \end{minipage}
\end{figure}

Consider component numbered $m$. The workers which don't straggle correspond to a set of continuous $c_m-d_m$ edges which is even and hence only a sum of $c_m-d_m$ gradients could be obtained corresponding to that cycle. An example demonstrating the fact is shown in Fig. \ref{fig:odd_edges} and Fig. \ref{fig:even_edges}. In Fig. \ref{fig:odd_edges}, if each worker transmits the sum of data-subsets assigned to it, we recover all the sum of gradients of data subsets spanned by it through transmissions from workers $W_1$ and $W_3$. However in Fig. \ref{fig:even_edges}, if each worker transmits the sum of data-subsets assigned to it, we cannot recover all the sum of gradients of all data subsets spanned by it through transmissions. We can at most recover the sum of gradients of data subsets $A_1$ and $A_2$ or $A_2$ and $A_3$ since the number of workers is even.

\remove{ 

Consider component numbered $m$. The workers which don't straggle correspond to a set of continuous $c_m-d_m$ edges which is even and hence only a sum of $c_m-d_m$ gradients could be obtained corresponding to that cycle. This is because we cannot recover the sum of all gradients spanned by a set of even consecutive non-straggling workers when each worker just transmits the sum of gradients assigned to it.

}

An example demonstrating the fact is shown in Fig. \ref{fig:odd_edges} and Fig. \ref{fig:even_edges}. In Fig. \ref{fig:odd_edges}, if each worker transmits the sum of data-subsets assigned to it, we recover all the sum of gradients of data subsets spanned by it through transmissions from workers $W_1$ and $W_3$. In Fig. \ref{fig:even_edges}, if each worker transmits the sum of data-subsets assigned to it, we can at most recover the sum of gradients of data subsets $A_1$ and $A_2$ or $A_2$ and $A_3$ since the number of workers is even. Similarly in component numbered $j$, there are exactly $c_j-d_j$ (which is even) workers corresponding to a set of continuous edges which don't straggle, hence only a sum of $c_j-d_j$ gradients could be obtained corresponding to that cycle.

\remove{Either the workers which don't straggle form the entire set of edges in the cycle in which case we obtain the sum of gradients corresponding to all the  data subsets in the cycle, or the workers which don't straggle form a continuous set of even edges potentially excluding one edge in the cycle in which case only a sum of $c_t-d_t$ gradients corresponding to the  data subsets in the cycle can be computed.}

For any other cycle numbered $i>m$ if no worker is straggled we can obtain the sum of gradients of all the  data subsets representing the nodes in the cycle i.e. a sum of gradients of $c_i-d_i=c_i$  data subsets. Thus, we could obtain a sum of gradients of exactly $\sum_{i =m}^{t} c_i-d_i= n-s = \beta-1$  data subsets which contradicts the requirement that the master should compute a sum of gradients of at least $\beta$  data subsets.

\remove{

Thus, exactly $s-1$ gradients cannot be computed as they are not present in any of the $n-s$ workers which returned. Now, consider the $m^{th}$ component. Here exactly $c_t-d_t$ workers corresponding to continuous edges in $m^{th}$ component don't straggle in this component. However $c_t-d_t$ is even and each worker returns the sum of the gradients computed by it as shown above. This would imply that sum of all $c_t-d_t+1$ gradients contained in the span of these set of $c_t-d_t$ non-stragglers, thus we won't have the master computing the sum of any set of $\beta$ gradients as there would be at least $s$ gradients not present in the sum computed at the master. 
}
\remove{{\bf Case-II:  The graph consists of disconnected components each of which need not be a cycle.} The proof of this case also is based on applying Claim \ref{claim:sum_grad} along with some additional arguments and the details are given in \cite{sarmasarkar2021gradient}.}

\textbf{\label{second_case}
Case II: Suppose the above graph is composed of $t$ distinct disconnected components. \remove{with no component having more than one cycle in it. Note that we even consider self-loops as cycles, thus implying no more than one self loop in any component}
}
 \remove{Like the previous example we divide the graph into various disjoint components}

\remove{
Note that every component would have at most the same number of edges as the number of nodes contained in it since it has exactly one cycle, however since the total number of nodes equals the total number of edges we can argue that each component must have exactly one cycle.
}

Suppose the edges of acyclic components is given by set $E_1$.We denote an order of removing edges in $E_1$ such that no new disconnected component is created at an step. Note that such an ordering can be ensured if we remove edges starting from a leaf node. Also observe that since no new disconnected component is created and only edges in acyclic components are removed, removal of $t$ edges would ensure at most $n-t$ nodes in its span since no new disconnected component is created in the process.

Now after these edges are removed, consider the largest set of edges (and self-loops) (denoted by set $E_2$) that can be removed so that each component has at least one cycle. Note that these edges would be removed in order such that no new disconnected component with one or more isolated edges is created in the process of removal of edges thus, ensuring at most $n-t$ nodes in the span of $n-t$ remaining edges in every step. Thus, after removal of $|E_1|+|E_2|$ edges, we would have distinct components with each component being a cycle.

\begin{figure}[!h]
    	\centering
    	\includegraphics[scale= 0.5]{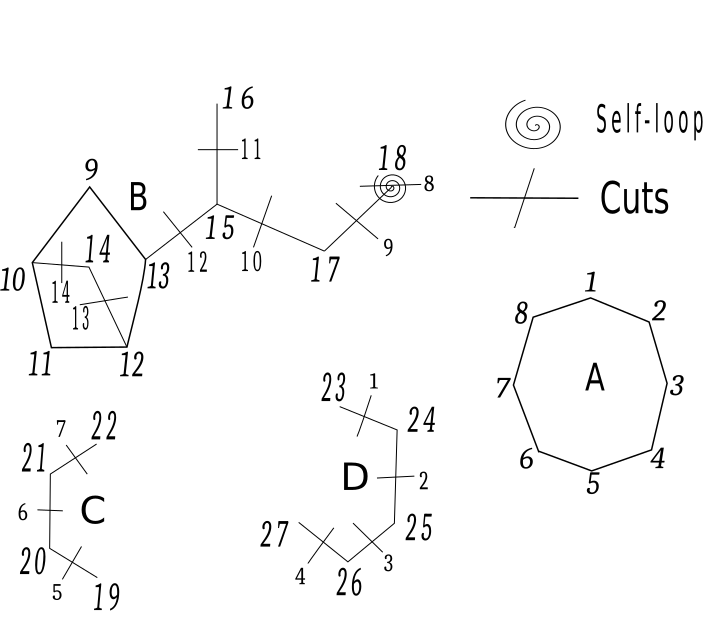}
    	\caption{Representation as data subsets as nodes with numbers denoting nodes and cuts denoting removal of edges}
    	\label{graph_representation_2}
\end{figure}

Note that cuts on edges Fig. \ref{graph_representation_2} denotes the edges in sets $E_1$ and $E_2$ in order described above with $n=28$. Also $E_1$ corresponds to edges numbered 1-7 and $E_2$ corresponds to edges numbered 8-14. Also note at no stage while removing edges in the order as numbered is a new disconnected component created with one or more isolated edges.

Thus, the maximum number of edges (or self-loops) that can be removed without reducing the number of cycles in the graph is denoted by $c_{max}=|E_1|+|E_2|$. Note that after removal of $|E_1|+|E_2|$ edges, we would have distinct components with each component being a cycle. Let us now consider three different conditions on $s$(maximum number of stragglers) and prove it in each case.

\textbf{
Case II-(a): $s\leq c_{max}= |E_1|+|E_2|$\\
}

Since the process of removing edges in each step ensures that exactly $n-t$ nodes lie in the span of remaining $n-t$ edges, we can just straggle the workers corresponding to the first $s$ edges in the process described above and argue that exactly $n-s=\beta-1$ data-subsets are accessible to the master, thus leading to a contradiction.  

Note that if we start straggling the workers corresponding to the edges in order ensure that no new disconnected component is created after the straggling workers are removed from the graph. Thus, removal of $v$ edges would ensure at most $n-v$ nodes in the span of remaining edges. This would follow from the fact that each component continues to have exactly one cycle. \remove{and removal of every corresponds to removal of at least one node from the span of remaining edges. }Such an ordered removal of edges can be done if we remove edges along a path starting from a leaf node to a cycle. 

Hence, if the total number of straggling workers is less than or equal to the maximum number of edges that can be removed without affecting any cycle, we can argue that no more than $n-s=\beta-1$ nodes can lie in its span, thus implying the master cannot compute the sum of $\beta$ gradients from the set of non-straggling workers.

\textbf{
Case II-(b): $s\geq c_{max}+2= |E_1|+|E_2|+2$\\
}

Under this constraint we first straggle all the edges corresponding to $E_1\cup E_2$ to obtain a similar structure of distinct cycles as the situation in Case \ref{first_case}. We now proceed in a very similar way as in Case \ref{first_case} and show that there exist a set of stragglers such that the master cannot compute the sum of any set of $\beta$ gradients. Note that, we require $s\geq c_{max}+2= |E_1|+|E_2|+2$ since the previous case also assumes at least 2 stragglers since $s=n-\beta+1\geq 2$ with $\beta \leq n-1$.

\textbf{
Case II-(c): $s = c_{max}+1= |E_1|+|E_2|+1$\\
}

Suppose there are more than 2 vertices in a cycle into which paths from leave nodes branch into. Well in this case, actually we can show that all workers corresponding to every edges in the cycle would have the same coefficient for both the  data subsets. Suppose $A_1,A_2,A_3$ be three consecutive nodes in the cycle and from each node there exists a cyclic path to a leaf node. Now we straggle the entire path to a leaf node from $A_2$, the edge $A_p-A_1$ (assuming $p$ nodes in the cycle) and edge $A_2-A_3$, the entire path from a leaf node to $A_3$ except the edge which connects it to the cycle. Using these set of stragglers we can show that the worker corresponding to the edge $A_1-A_2$ has same coefficient for both the data subsets. Similarly, we can argue that the workers corresponding to all edges have the same coefficient for both the workers and a similar selection of stragglers will show that no sum of $\beta$ gradients exists in the linear span of non-straggling workers.

Suppose there are at most two vertices in the cycle into which paths from leaf nodes merge into. In this case, there might be at most two edges in the cycle the workers corresponding to which may not have the same coefficient for the gradients of the data subsets assigned to them. 

Suppose there exists only a path to node $A_1$ in the cycle to a leaf node of length $c_{max}$, thus in this example we can show that all workers except those corresponding to edges $A_1-A_2$ and $A_n-A_1$ have the same coefficient for both the gradients of the data subsets assigned to it. 

Now if $p$(the edges in cycle) is odd, straggle the workers corresponding to the edges in the path from $A_1$ to leaf node and the edge $A_1-A_2$. We can show that since an even number of consecutive edges remain in the component after straggling the edges we cannot have a sum of all the gradients of the data subsets contained in the span of the edges corresponding to non-straggling workers. Since $\beta$ data subsets remain in the span after the selection of straggling workers, we cannot have any sum of $\beta$ gradients in its span.

Suppose $p$ is even, straggle the workers corresponding to the edges in the path from $A_1$ to leaf node and the edge $A_1-A_2$. Now the number of edges remaining in this component is odd which would imply that there must exist some other non-straggling cyclic component with odd number of edges as the total number of non-straggling edges is $n-s=\beta-1$ which is even. Also there must exist a cyclic component after removal of $c_{max}$ edges which has an odd number of edges. Thus instead of straggling node $A_1-A_2$, straggle a node in another cyclic component with odd number of edges and the same argument as above follows. Similarly, we can argue for the other cases too if two nodes in a cycle have paths from leaf nodes branching into it.

\section{Proof of theorem \ref{lower_bound2_low_comp}}
{\label{lower_bound2_low_comp_section}}
    We prove by contradiction. Suppose we have $l=\frac{2}{n}$ i.e. every worker transmits and computes a linear combination of gradients of at most two data subsets. We represent every data subsets as a node in the graph. As, every worker can be assigned at most 2 data subsets, we denote it as an edge if it indeed transmits a linear combination of two data subsets, else we represent it a self-loop around a node corresponding to the data subset assigned to it.

Suppose the graph is represented in $t$ distinct components and suppose $u$ components have no cycle in it. There might be some components with multiple cycles in it too. Now we define an order of straggling workers. First start with those components which don't have any cycle and straggle the workers corresponding to those edges which start from a leaf edge and straggle workers continuously along a path. Note that this process would ensure that no new component is created when the edges corresponding to straggling workers are removed from the graph, thus at most $n-t$ nodes in the span when $t$ edges are removed. Now consider the maximum number of edges that can be removed (corresponding workers straggled) such that each cyclic component continues to have at least one cycle. Note that we straggle these edges in order such that no new component is created in the process of removing edges from the graph. This would ensure that there is at most $n-t$ nodes in the span of remaining edges when $t$ workers are straggled (removed) from the graph. At the end, we would have only cyclic components remaining. Straggle the edges in each cycle continuously till no edge in a cycle is left and then start with the next cycle. 

We stop when there is no worker is left. Note that in this process of straggling, when $n-v$ edges corresponding to workers are remaining we can have atmost $n-v+1$ vertices or data subsets being spanned. This can be argued from the fact that unless we break a new cycle a data subset is always removed from the span whenever a worker is straggled and no distinct component of the graph is being created in the process of straggling (removing) workers. Thus straggling of $s$ workers under the above mentioned process would ensure at most $n-s+1 \leq \beta-1$ data subsets being accessible at the master implying a contradiction.

\remove{

After all the workers corresponding to edges in non-cyclic components are straggled and removed from the graph, straggle those workers corresponding to edges in cyclic components such that each component has at least one cycle when with no new disconnected component being created when the edges corresponding to straggled workers are removed.

which are not part of any cycle again starting from a leaf edge and straggle continuously along a path. Now choose a component with more than one cycle and straggle workers continuously along a path after one cycle is reduced. Workers are straggled along a new path only when no more edges can be removed without a breaking a new cycle. Note that this process would ensure a connected component remains connected even while straggling workers are removed as edges or self-loops from the graph.

At the end, we would have only cyclic components remaining. Straggle the edges in each cycle continuously till no edge in a cycle is left and then start with the next cycle. We stop when there is no worker is left.

}

%% file: proof_interim_points.tex
\section{Proof of Theorem \ref{high_comm_cyclic_middle}}{\label{high_comm_cyclic_appendix}}

 First, we assume that the indices of the non-straggled workers are consecutive, w.l.o.g, $1,2,\ldots, n-s$. Consider all data-partitions not assigned to any of these workers whose $i^{th}$ element in the corresponding list is the largest, suppose we denote such a list by $\{c_1,c_2,\ldots c_y\}$. Thus, $c_i \leq (n-\gamma_i+1)$, $(c_{(j \mod n)+1)}-c_j) \geq \gamma_j$ for $j \in [y]$ and $c_{i+1}\geq n-s+1$. Thus, the number of such lists with the largest element being the $i^{th}$ element is given by ${{s-(\gamma_i)-\sum_{i=1}^{y-1}\gamma_i+y \choose y}}={{s-\delta+y \choose y}}$ as $\delta= \sum_{i=1}^{y}\gamma_i$. Thus, the total number of data-partitions not assigned to any of the consecutive $n-s$ workers is given by $y.{{s-\delta+y \choose y}}$. Thus, the number of gradients received at the master when consecutive $s$ workers straggle is given by $n{{n-\delta+y-1 \choose y-1}}-y{{s-\delta+y \choose y}}$.

Suppose the straggling workers are not consecutive and can be split into $(t+1)$ disjoint consecutive groups of indices denoted by $I_1,I_2,\ldots,I_{t+1}$ which are increasingly ordered. Now consider any data-partition which is not assigned to any of the non-straggling $n-s$ workers and we consider lists corresponding to all such data-partitions such that the $i^{th}$ element in the list being the smallest. Note that all the elements in the list corresponding to the data-partition not assigned to any of the $n-s$ non-straggling workers would be contained in either $I_1,I_2,\ldots,I_{t+1}$. 

Now we construct a list denoted by $J$ of length $s$ by appending all elements in $I_1,I_2,\ldots,I_{t+1}$ in order. Now consider all the elements in this list which are the elements of the list  $K=[c_1,c_2,\ldots,c_y]$ denoting the data-partition not assigned to any of non-straggling workers which precisely contains $y$ elements. Note that the elements in the list would be ordered as $c_i<c_{i+1}<\ldots<c_y<c_1<\ldots <c_{i-1}$. 

\remove{
 W.L.O.G, we assume that that the elements in $I_1$ are numbered from $\{1,2,\ldots,z_1\}$ and the elements in $I_2$ are numbered from $\{z_2,\ldots,z_3\}$ and so on till $\{z_{2(t-1)},\ldots,z_{2t-1}\}$. Note that $1\leq z_1 <z_2 \leq z_3 < z_4 \leq z_5 ...<z_{2((t-1)}\leq z_{2t-1}$.
We can argue that the indices of these elements in $J$ would also be separated by $\gamma_i,\gamma_{i+1},\ldots,\gamma_y,\gamma_1,\ldots$ and the index of the element $c_{i-1}$ in $J$ would be at most $s-\gamma_{i-1}+1$. We argue by contradiction.
}

Consider any two elements $c_t$ and $c_{t+1}$ and suppose they belong in the same list $I_u$, in which case their indices in $J$ must differ by at least $\gamma_t$ to satisfy the condition that $c_{t+1}-c_t \geq \gamma_t$. Suppose they belong in two different
lists $I_u$ and $I_v$ with $v>u$, in which case there must be at least $\gamma_t-1$ elements larger than $c_t$ in list $I_u$. This is because, workers numbered $c_t,c_t+1,\ldots,c_t+\gamma_t-1$ would be assigned the data subset corresponding to list $K=[c_1,c_2,\ldots,c_y]$ each of which must be a straggling worker as this data subset is not assigned to any of the non-straggling workers. Thus, the elements from $c_t,c_t+1,\ldots,c_t+\gamma_t-1$ would be present in list $I_u$, hence there would be at least $\gamma_t-1$ elements between $c_t$ and $c_{t+1}$ as all elements of list $I_u$ have lower index in $J$ than those elements of list $I_v$ as $v>u$. Similarly, there would be at least $\gamma_{i-1}-1$ with index higher than $c_{i-1}$ in list $J$.

Thus, the indices of these elements in $J$ would also differ by $\gamma_i,\gamma_{i+1},\ldots,\gamma_y,\gamma_1,\ldots$ and the index of the element $c_{i-1}$ in $J$ would be at most $s-\gamma_{i-1}+1$. Also note that the indices of these elements in $J$ would be different for every distinct list $[c_1,c_2,\ldots,c_y]$. Thus, the number of such lists having the $i^{th}$ element as the smallest index would be bounded by ${{s-\delta+y \choose y}}$ (as the setting is similar to the case of $n-s$ consecutive non-straggling workers), thus the total number of such lists is bounded by $y{{s-\delta+y \choose y}}$. Hence at most $y{{s-\delta+y \choose y}}$ data subsets would not be assigned to any of $n-s$ non-straggling workers, hence this scheme would be $(\alpha,s)$ feasible for $\alpha \leq 1- \frac{y{{s-\delta+y \choose y}}}{n{{n-\delta+y-1 \choose y-1}}}$.

Let us now compute the number of data-partitions not assigned to one worker. Using a similar technique as used previously, we argue that $y{{n-1-\delta+y \choose y}}$ data subsets would not be assigned to that worker, thus the communication cost incurred by this scheme is $(n{{n-\delta+y-1 \choose y-1}}-y.{{n-1-\delta+y \choose y}})=({{n-\delta+y-1 \choose y}})\frac{y\delta}{n-\delta}=({{n-\delta+y-1 \choose y-1}})\delta$.

%% file: main.bbl
\begin{thebibliography}{10}
\providecommand{\url}[1]{#1}
\csname url@samestyle\endcsname
\providecommand{\newblock}{\relax}
\providecommand{\bibinfo}[2]{#2}
\providecommand{\BIBentrySTDinterwordspacing}{\spaceskip=0pt\relax}
\providecommand{\BIBentryALTinterwordstretchfactor}{4}
\providecommand{\BIBentryALTinterwordspacing}{\spaceskip=\fontdimen2\font plus
\BIBentryALTinterwordstretchfactor\fontdimen3\font minus
  \fontdimen4\font\relax}
\providecommand{\BIBforeignlanguage}[2]{{%
\expandafter\ifx\csname l@#1\endcsname\relax
\typeout{** WARNING: IEEEtran.bst: No hyphenation pattern has been}%
\typeout{** loaded for the language `#1'. Using the pattern for}%
\typeout{** the default language instead.}%
\else
\language=\csname l@#1\endcsname
\fi
#2}}
\providecommand{\BIBdecl}{\relax}
\BIBdecl

\bibitem{li2020coded}
S.~Li and S.~Avestimehr, ``Coded computing,'' \emph{Foundations and
  Trends{\textregistered} in Communications and Information Theory}, vol.~17,
  no.~1, 2020.

\bibitem{dutta2016short}
S.~Dutta, V.~Cadambe, and P.~Grover, ``Short-dot: computing large linear
  transforms distributedly using coded short dot products,'' in
  \emph{Proceedings of the 30th International Conference on Neural Information
  Processing Systems}, 2016, pp. 2100--2108.

\bibitem{lee2017speeding}
K.~Lee, M.~Lam, R.~Pedarsani, D.~Papailiopoulos, and K.~Ramchandran, ``Speeding
  up distributed machine learning using codes,'' \emph{IEEE Transactions on
  Information Theory}, vol.~64, no.~3, pp. 1514--1529, 2017.

\bibitem{yu2017polynomial}
Q.~Yu, M.~A. Maddah-Ali, and A.~S. Avestimehr, ``Polynomial codes: an optimal
  design for high-dimensional coded matrix multiplication,'' in
  \emph{Proceedings of the 31st International Conference on Neural Information
  Processing Systems}, 2017, pp. 4406--4416.

\bibitem{tandon2017gradient}
R.~Tandon, Q.~Lei, A.~G. Dimakis, and N.~Karampatziakis, ``Gradient coding:
  Avoiding stragglers in distributed learning,'' in \emph{International
  Conference on Machine Learning}.\hskip 1em plus 0.5em minus 0.4em\relax PMLR,
  2017, pp. 3368--3376.

\bibitem{yu2019lagrange}
Q.~Yu, S.~Li, N.~Raviv, S.~M.~M. Kalan, M.~Soltanolkotabi, and S.~A.
  Avestimehr, ``Lagrange coded computing: Optimal design for resiliency,
  security, and privacy,'' in \emph{The 22nd International Conference on
  Artificial Intelligence and Statistics}.\hskip 1em plus 0.5em minus
  0.4em\relax PMLR, 2019, pp. 1215--1225.

\bibitem{8006960}
S.~{Dutta}, V.~{Cadambe}, and P.~{Grover}, ``Coded convolution for parallel and
  distributed computing within a deadline,'' in \emph{2017 IEEE International
  Symposium on Information Theory (ISIT)}, 2017, pp. 2403--2407.

\bibitem{li2017fundamental}
S.~Li, M.~A. Maddah-Ali, Q.~Yu, and A.~S. Avestimehr, ``A fundamental tradeoff
  between computation and communication in distributed computing,'' \emph{IEEE
  Transactions on Information Theory}, vol.~64, no.~1, pp. 109--128, 2017.

\bibitem{raviv2020gradient}
N.~Raviv, I.~Tamo, R.~Tandon, and A.~G. Dimakis, ``Gradient coding from cyclic
  mds codes and expander graphs,'' \emph{IEEE Transactions on Information
  Theory}, vol.~66, no.~12, pp. 7475--7489, 2020.

\bibitem{halbawi2018improving}
W.~Halbawi, N.~Azizan, F.~Salehi, and B.~Hassibi, ``Improving distributed
  gradient descent using reed-solomon codes,'' in \emph{2018 IEEE International
  Symposium on Information Theory (ISIT)}.\hskip 1em plus 0.5em minus
  0.4em\relax IEEE, 2018, pp. 2027--2031.

\bibitem{ye2018communication}
M.~Ye and E.~Abbe, ``Communication-computation efficient gradient coding,'' in
  \emph{International Conference on Machine Learning}.\hskip 1em plus 0.5em
  minus 0.4em\relax PMLR, 2018, pp. 5610--5619.

\bibitem{ozfatura2019speeding}
E.~Ozfatura, D.~G{\"u}nd{\"u}z, and S.~Ulukus, ``Speeding up distributed
  gradient descent by utilizing non-persistent stragglers,'' in \emph{2019 IEEE
  International Symposium on Information Theory (ISIT)}.\hskip 1em plus 0.5em
  minus 0.4em\relax IEEE, 2019, pp. 2729--2733.

\bibitem{ozfatura2020straggler}
E.~Ozfatura, S.~Ulukus, and D.~G{\"u}nd{\"u}z, ``Straggler-aware distributed
  learning: Communication--computation latency trade-off,'' \emph{Entropy},
  vol.~22, no.~5, p. 544, 2020.

\bibitem{wang2019heterogeneity}
H.~Wang, S.~Guo, B.~Tang, R.~Li, and C.~Li, ``Heterogeneity-aware gradient
  coding for straggler tolerance,'' in \emph{2019 IEEE 39th International
  Conference on Distributed Computing Systems (ICDCS)}.\hskip 1em plus 0.5em
  minus 0.4em\relax IEEE, 2019, pp. 555--564.

\bibitem{9614153}
K.~Wan, H.~Sun, M.~Ji, and G.~Caire, ``Distributed linearly separable
  computation,'' \emph{IEEE Transactions on Information Theory}, vol.~68,
  no.~2, pp. 1259--1278, 2022.

\bibitem{dutta2018slow}
S.~Dutta, G.~Joshi, S.~Ghosh, P.~Dube, and P.~Nagpurkar, ``Slow and stale
  gradients can win the race: Error-runtime trade-offs in distributed sgd,'' in
  \emph{International Conference on Artificial Intelligence and
  Statistics}.\hskip 1em plus 0.5em minus 0.4em\relax PMLR, 2018, pp. 803--812.

\bibitem{chen2017revisiting}
J.~Chen, X.~Pan, R.~Monga, S.~Bengio, and R.~Jozefowicz, ``Revisiting
  distributed synchronous sgd,'' \emph{arXiv preprint arXiv:1604.00981}, 2016.

\bibitem{bottou2010large}
L.~Bottou, ``Large-scale machine learning with stochastic gradient descent,''
  in \emph{Proceedings of COMPSTAT'2010}.\hskip 1em plus 0.5em minus
  0.4em\relax Springer, 2010, pp. 177--186.

\bibitem{wang2019erasurehead}
H.~Wang, Z.~Charles, and D.~Papailiopoulos, ``Erasurehead: Distributed gradient
  descent without delays using approximate gradient coding,'' \emph{arXiv
  preprint arXiv:1901.09671}, 2019.

\bibitem{9081964}
R.~Bitar, M.~Wootters, and S.~El~Rouayheb, ``Stochastic gradient coding for
  straggler mitigation in distributed learning,'' \emph{IEEE Journal on
  Selected Areas in Information Theory}, vol.~1, no.~1, pp. 277--291, 2020.

\bibitem{maity2019robust}
R.~K. Maity, A.~S. Rawa, and A.~Mazumdar, ``Robust gradient descent via moment
  encoding and ldpc codes,'' in \emph{2019 IEEE International Symposium on
  Information Theory (ISIT)}.\hskip 1em plus 0.5em minus 0.4em\relax IEEE,
  2019, pp. 2734--2738.

\bibitem{wang2019fundamental}
S.~Wang, J.~Liu, and N.~Shroff, ``Fundamental limits of approximate gradient
  coding,'' \emph{Proceedings of the ACM on Measurement and Analysis of
  Computing Systems}, vol.~3, no.~3, pp. 1--22, 2019.

\bibitem{8849690}
S.~{Kadhe}, O.~O. {Koyluoglu}, and K.~{Ramchandran}, ``Gradient coding based on
  block designs for mitigating adversarial stragglers,'' in \emph{2019 IEEE
  International Symposium on Information Theory (ISIT)}, 2019, pp. 2813--2817.

\bibitem{charles2017approximate}
Z.~Charles, D.~Papailiopoulos, and J.~Ellenberg, ``Approximate gradient coding
  via sparse random graphs,'' \emph{arXiv preprint arXiv:1711.06771}, 2017.

\bibitem{DBLP:journals/corr/abs-2002-11005}
\BIBentryALTinterwordspacing
S.~K. Hanna, R.~Bitar, P.~Parag, V.~Dasari, and S.~E. Rouayheb, ``Adaptive
  distributed stochastic gradient descent for minimizing delay in the presence
  of stragglers,'' \emph{CoRR}, vol. abs/2002.11005, 2020. [Online]. Available:
  \url{https://arxiv.org/abs/2002.11005}
\BIBentrySTDinterwordspacing

\bibitem{ozfatura2019distributed}
E.~Ozfatura, S.~Ulukus, and D.~G{\"u}nd{\"u}z, ``Distributed gradient descent
  with coded partial gradient computations,'' in \emph{ICASSP 2019-2019 IEEE
  International Conference on Acoustics, Speech and Signal Processing
  (ICASSP)}.\hskip 1em plus 0.5em minus 0.4em\relax IEEE, 2019, pp. 3492--3496.

\bibitem{9641837}
E.~Ozfatura, S.~Ulukus, and D.~Gündüz, ``Coded distributed computing with
  partial recovery,'' \emph{IEEE Transactions on Information Theory}, vol.~68,
  no.~3, pp. 1945--1959, 2022.

\bibitem{10.5555-1202540}
C.~J. Colbourn and J.~H. Dinitz, \emph{Handbook of Combinatorial Designs,
  Second Edition (Discrete Mathematics and Its Applications)}.\hskip 1em plus
  0.5em minus 0.4em\relax Chapman \& Hall/CRC, 2006.

\bibitem{9174030}
P.~Peng, E.~Soljanin, and P.~Whiting, ``Diversity vs. parallelism in
  distributed computing with redundancy,'' in \emph{2020 IEEE International
  Symposium on Information Theory (ISIT)}, 2020, pp. 257--262.

\bibitem{MNIST_data}
C.~C. Yann~LeCun, ``Mnist data set,'' \url{http://yann.lecun.com/exdb/mnist/}.

\end{thebibliography}
